\documentclass[10pt,aps,pra,superscriptaddress,twocolumn]{revtex4-2}
\usepackage[english]{babel}
\usepackage{amsmath}
\usepackage{amsthm}
\usepackage{amssymb}
\usepackage{mathrsfs}
\usepackage{booktabs}
\usepackage{color}

\usepackage[ruled,vlined]{algorithm2e}
\usepackage{cases}
\usepackage{multirow}
\usepackage{xcolor}
\usepackage{physics}

\usepackage{mathtools}
\usepackage{esint}
\usepackage{natbib}

\usepackage{hyperref}
\usepackage[capitalise]{cleveref}

\usepackage{tikz}
\usepackage[normalem]{ulem}

\newcommand{\efimovspecs}{2 $\times$ Intel(R) Xeon(R) Gold 6240}

\begin{abstract}
    The evolution of a quantum system under time-dependent driving exhibits phenomena that are absent in its stationary counterpart. However, the high dimensionality and non-commutative nature of quantum dynamics make this a challenging problem. 
    The Magnus expansion provides an analytic framework to approximate the effective dynamics on short time-scales, but computing high-order terms with existing methods is computationally expensive.
    We introduce a scalable approach that reduces the computational effort to depend only on the degrees of freedom defining the \text{time-dependent} control function. 
    We focus specifically on Hamiltonians consisting of a constant drift term and a controllable term. Our method provides a polynomial expression for the Magnus expansion which can be evaluated several orders of magnitude faster than previous techniques, enabling broad applications in the realms of quantum simulation and quantum optimal control.  We showcase an application of the method by designing control pulses for the 5-qubit phase gate on a neutral-atom platform utilizing Rydberg atoms. 
\end{abstract}

\begin{document}
\title{ \texorpdfstring{Lie algebra-assisted quantum simulation and quantum optimal control\\
via high-order Magnus expansions}{Lie algebra-assisted quantum simulation and quantum optimal control
via high-order Magnus expansions}}

\author{R.F. \surname{dos Santos}}
\affiliation{
Department of Applied Physics, Eindhoven University of Technology, P. O. Box 513, 5600 MB Eindhoven, The Netherlands}
\author{S.J.J.M.F. \surname{Kokkelmans}}
\affiliation{
Department of Applied Physics, Eindhoven University of Technology, P. O. Box 513, 5600 MB Eindhoven, The Netherlands}
\date{\today}
\maketitle

\newcommand{\circbfix}{%
  \tikz[baseline=(c.base)]{
    \node[circle, fill=black, inner sep=0pt, minimum size=0.8ex] (c) {};
  }%
}

\tikzset{
  smallpoint/.style={
    inner sep=0pt,
    outer sep=0pt,
    path picture={\circbfix}
  }
}
\newcommand{\tafix}{%
\raisebox{-10pt}{%
\begin{tikzpicture}[
  grow=up,
  level distance=3mm,
  sibling distance=4mm,
  every node/.style={inner sep=0pt, outer sep=0pt},
  level 1/.style={sibling distance=6mm, level distance=3mm},
  level 2/.style={sibling distance=6mm, level distance=4mm},
  level 3/.style={sibling distance=4mm, level distance=2mm}
]
  \coordinate[grow=up]
    child { node (t1) {} }
    child { child { node (t2) {} } };
  \node[above] at (t1) {$\tau_2$};
  \node[above] at (t2) {$\tau_1$};
\end{tikzpicture}%
}}

\newcommand{\tbfix}{%
\raisebox{-5pt}{%
\begin{tikzpicture}[
  grow=up,
  level distance=3mm,
  sibling distance=4mm,
  every node/.style={inner sep=0pt, outer sep=0pt},
  level 1/.style={sibling distance=4mm, level distance=2mm},
  level 2/.style={sibling distance=4mm, level distance=2mm},
  level 3/.style={sibling distance=4mm, level distance=2mm}
]
  \coordinate[grow=up]
    child { node[smallpoint] {\circbfix} }
    child { child { node[smallpoint] {\circbfix} } };
\end{tikzpicture}%
}
\
}

\newcommand{\tbifix}{%
\raisebox{-10pt}{%
\begin{tikzpicture}[
  grow=up,
  level distance=3mm,
  sibling distance=4mm,
  every node/.style={inner sep=0pt, outer sep=0pt},
  level 1/.style={sibling distance=4mm, level distance=2mm},
  level 2/.style={sibling distance=4mm, level distance=2mm},
  level 3/.style={sibling distance=4mm, level distance=2mm}
]
  \coordinate[grow=up]
    child {
      child { node[smallpoint] {\circbfix} }
      child { child { node[smallpoint] {\circbfix} } }
    };
\end{tikzpicture}%
}
\quad 
}

\newcommand{\tcfix}{%
\
\raisebox{-10pt}{%
\begin{tikzpicture}[
  grow=up,
  level distance=3mm,
  sibling distance=4mm,
  every node/.style={inner sep=0pt, outer sep=0pt},
  level 1/.style={sibling distance=4mm, level distance=2mm},
  level 2/.style={sibling distance=4mm, level distance=2mm},
  level 3/.style={sibling distance=4mm, level distance=2mm}
]
  \coordinate[grow=up]
    child { node[smallpoint] {\circbfix} }
    child {
      child {
        child { node[smallpoint] {\circbfix} }
        child {
          child { node[smallpoint] {\circbfix} }
        }
      }
    };
\end{tikzpicture}%
}
\ \ 
}

\newcommand{\tdfix}{%
\raisebox{-10pt}{%
\
\begin{tikzpicture}[
  grow=up,
  level distance=3mm,
  sibling distance=4mm,
  every node/.style={inner sep=0pt, outer sep=0pt},
  level 1/.style={sibling distance=4mm, level distance=2mm},
  level 2/.style={sibling distance=4mm, level distance=2mm},
  level 3/.style={sibling distance=4mm, level distance=2mm}
]
  \coordinate[grow=up]
    child {
      child { node[smallpoint] {\circbfix} }
      child {
        child { node[smallpoint] {\circbfix} }
      }
    }
    child { child { node[smallpoint] {\circbfix} } };
\end{tikzpicture}%
}
\quad
}

\newcommand{\tdifix}{%
\raisebox{-10pt}{%
\
\begin{tikzpicture}[
  grow=up,
  level distance=3mm,
  sibling distance=4mm,
  every node/.style={inner sep=0pt, outer sep=0pt},
  level 1/.style={sibling distance=4mm, level distance=2mm},
  level 2/.style={sibling distance=4mm, level distance=2mm},
  level 3/.style={sibling distance=4mm, level distance=2mm}
]
  \coordinate[grow=up]
    child {
      child {
        child { node[smallpoint] {\circbfix} }
        child {
          child { node[smallpoint] {\circbfix} }
        }
      }
      child {
        child { node[smallpoint] {\circbfix} }
      }
    };
\end{tikzpicture}%
}
\quad
}

\newcommand{\tefix}{%
\ 
\raisebox{-15pt}{%
\begin{tikzpicture}[
  grow=up,
  level distance=3mm,
  sibling distance=4mm,
  every node/.style={inner sep=0pt, outer sep=0pt},
  level 1/.style={sibling distance=4mm, level distance=2mm},
  level 2/.style={sibling distance=4mm, level distance=2mm},
  level 3/.style={sibling distance=4mm, level distance=2mm}
]
  \coordinate[grow=up]
    child { node[smallpoint] {\circbfix} }
    child {
      child {
        child {
          child { node[smallpoint] {\circbfix} }
          child { child { node[smallpoint] {\circbfix} } }
        }
        child { child { node[smallpoint] {\circbfix} } }
      }
    };
\end{tikzpicture}%
}
\quad
}

\newcommand{\teifix}{%
\raisebox{-15pt}{%
\begin{tikzpicture}[
  grow=up,
  level distance=3mm,
  sibling distance=4mm,
  every node/.style={inner sep=0pt, outer sep=0pt},
  level 1/.style={sibling distance=4mm, level distance=2mm},
  level 2/.style={sibling distance=4mm, level distance=2mm},
  level 3/.style={sibling distance=4mm, level distance=2mm}
]
  \coordinate[grow=up]
    child {
      child { node {\circbfix} } 
      child {                     
        child {
          child {
            child { node {\circbfix} }           
            child { child { node {\circbfix} } } 
          }
          child { child { node {\circbfix} } }   
        }
      }
    };
\end{tikzpicture}%
}}

\newcommand{\tffix}{%
\raisebox{-15pt}{%
\begin{tikzpicture}[
  grow=up,
  level distance=3mm,
  sibling distance=4mm,
  every node/.style={inner sep=0pt, outer sep=0pt},
  level 1/.style={sibling distance=4mm, level distance=2mm},
  level 2/.style={sibling distance=4mm, level distance=2mm},
  level 3/.style={sibling distance=4mm, level distance=2mm}
]
  \coordinate[grow=up]
    child { node[smallpoint] {\circbfix} }
    child {
      child {
        child { node[smallpoint] {\circbfix} }
        child {
          child {
            child {
              child { node[smallpoint] {\circbfix} }
              child { child { node[smallpoint] {\circbfix} } }
            }
          }
        }
      }
    };
\end{tikzpicture}%
}
\ \
}

\newcommand{\tgfix}{%
\
\raisebox{-15pt}{%
\begin{tikzpicture}[
  grow=up,
  level distance=3mm,
  sibling distance=4mm,
  every node/.style={inner sep=0pt, outer sep=0pt},
  level 1/.style={sibling distance=6mm, level distance=3mm},
  level 2/.style={sibling distance=4mm, level distance=2mm},
  level 3/.style={sibling distance=4mm, level distance=2mm}
]
  \coordinate[grow=up]
    child {
      child {
        child { node[smallpoint] {\circbfix} }
        child { child { node[smallpoint] {\circbfix} } }
      }
    }
    child {
      child {
        child { node[smallpoint] {\circbfix} }
        child { child { node[smallpoint] {\circbfix} } }
      }
    };
\end{tikzpicture}%
}
\ \quad
}

\newcommand{\tifix}{%
\
\raisebox{-15pt}{%
\begin{tikzpicture}[
  grow=up,
  level distance=3mm,
  sibling distance=4mm,
  every node/.style={inner sep=0pt, outer sep=0pt},
  level 1/.style={sibling distance=4mm, level distance=2mm},
  level 2/.style={sibling distance=4mm, level distance=2mm},
  level 3/.style={sibling distance=4mm, level distance=2mm}
]
  \coordinate[grow=up]
    child {
      child { node[smallpoint] {\circbfix} }
      child {
        child {
          child {
            child { node[smallpoint] {\circbfix} }
            child { child { node[smallpoint] {\circbfix} } }
          }
        }
      }
    }
    child {
      child { node[smallpoint] {\circbfix} }
    };
\end{tikzpicture}%
}
\, \quad
}

\newcommand{\tjfix}{%
\
\raisebox{-10pt}{%
\begin{tikzpicture}[
  grow=up,
  level distance=3mm,
  sibling distance=4mm,
  every node/.style={inner sep=0pt, outer sep=0pt},
  level 1/.style={sibling distance=4mm, level distance=2mm},
  level 2/.style={sibling distance=4mm, level distance=2mm},
  level 3/.style={sibling distance=4mm, level distance=2mm}
]
  \coordinate[grow=up]
    child {
      child {
        child { node[smallpoint] {\circbfix} }
        child {
          child { node[smallpoint] {\circbfix} }
        }
      }
      child {
        child { node[smallpoint] {\circbfix} }
      }
    }
    child {
      child { node[smallpoint] {\circbfix} }
    };
\end{tikzpicture}%
}
\ \quad
}

\newcommand{\tuniquefix}{
\raisebox{-10pt}{
\begin{tikzpicture}[level distance=8mm,sibling distance=4mm, grow=up,
  every node/.style={inner sep=2pt, outer sep=0pt},
  level 1/.style={sibling distance=6mm, level distance=3mm},
  level 2/.style={sibling distance=6mm, level distance=3mm},
  level 3/.style={sibling distance=6mm, level distance=3mm}]
\coordinate[grow=up]
  child {
    child { 
      child {
        child { 
            node[circle, draw, fill=black, inner sep=1.5pt] {} 
            edge from parent [solid] 
        }
        child { 
          child { node (ts) {} edge from parent [solid] }
          edge from parent [solid]
        } 
        edge from parent [dotted]
      }    
      child { 
        child { node (t3) {} }
      }
    }    
    child { 
      child { node (t2) {} }
    }
  }
  child { 
    child { node (t1) {} }
  };
\node[above] at (ts) {$\tau_s$};
\node[above] at (t3) {$\tau_3$};
\node[above] at (t1) {$\tau_1$};
\node[above] at (t2) {$\tau_2$};
\end{tikzpicture}

}}

\section{Introduction}



Quantum dynamics underpins the development of emerging technologies, including quantum computation~\cite{PhysRevResearch.3.043212,Qiang2016}, sensing~\cite{Montenegro2025,Aslam2023,Zou2025}, and communication~\cite{Ramya2025}, as well as serving as a test-bed for fundamental physics~\cite{Bauer2023,vanDerMeer2022,GonzalezCuadra2024,PhysRevLett.133.111901}. Devices exploiting quantum phenomena are now beginning to appear, with quantum computers being a prominent example. These versatile platforms enable experimental validation of quantum information theory, encompassing topics such as quantum control~\cite{ansel2024introduction,mahesh2022quantum,chen2023quantum}, entanglement generation~\cite{su2023characterization}, ground-state preparation~\cite{dong2022ground,maskara2023programmablesimulationsmoleculesmaterials}, and quantum error correction~\cite{demo-lowoverhead-qldpc-32q,qec-google2024,PhysRevA.111.022433,Bluvstein2023}. Yet, in the current noisy intermediate-scale quantum (NISQ) era, available qubits remain scarce and highly susceptible to environmental noise, limiting practical advantages over classical approaches~\cite{Preskill2018,Chen2023}. Consequently, the design of efficient quantum operations from inherently complex quantum dynamics is critical for advancing quantum computing and overcoming the challenges of this era.



Neutral atom quantum computing platforms are especially interesting as a toolbox for studying complex quantum dynamics~\cite{Saffman2016,Henriet2020quantumcomputing,Morgado2021QuantumQubits,Scholl2021,r54t-myhc}. 
Since the interatomic interactions have a strong dependence on distance, different geometries serve as test-beds to simulate quantum materials and study the associated properties~\cite{maskara2023programmablesimulationsmoleculesmaterials,Scholl2021,r54t-myhc}. Coherent re-arrangeability through optical means is another prominent feature, allowing for flexible quantum circuits and quantum error correction demonstrations~\cite{Evered2023HighFidelity,Bluvstein2023}.


Harnessing the intricate dynamics of quantum systems demands operations that are both precise and coherence-preserving. Fundamental tasks in quantum computing--such as state preparation and gate synthesis--rely on engineered control pulses to shape the underlying physical evolution. Yet, the non-commutative nature of quantum dynamics often precludes closed-form analytical solutions, making simulation and optimization necessary for designing and understanding these operations on real devices. 
Lie groups and algebras have been found to be effective tools for describing and studying the non-commutative nature of quantum dynamics~\cite{dirr2008lie,gsim,Wei1963,Hall2015LieRepresentations,Wiersema2024,PRXQuantum.6.010201,Ragone2024ACircuits,ansel2024introduction}.

\begin{figure*}[t]
    \centering
    \includegraphics[width=0.8\linewidth]{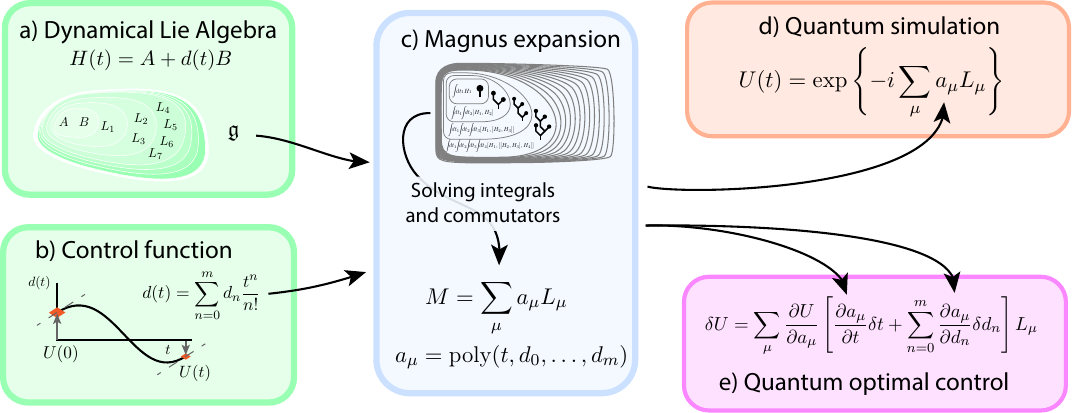}
    \caption{In this work we propose a method to efficiently evaluate the Magnus expansion. (a) Diagram of the generation of the the dynamical Lie algebra that stems from the single-control Hamiltonian terms $A,B$. (b) The control function is represented as a polynomial with maximum degree $m$, which allow us to solve high-order integrals analytically. (c) The high-order integrals and commutators from the Magnus expansion are simplified into a polynomial expression. (d)  This enables high-order simulation of continuous-controlled quantum systems. (e) The analytically-differentiable expression enables quantum optimal control applications.}
    \label{fig:graphical-abstract}
\end{figure*}

The Magnus expansion (ME) is an exponential representation of the solution operator of the quantum system that emerges from the study of Lie algebras \cite{magnusapplications,Milfeld1983-floquetMagnus}. It is a series expansion that approximates the solution operator of the time-dependent quantum system. This solution operator is represented as a constant Hamiltonian, often referred to as the effective Hamiltonian, and it evolves the quantum system the same way as if evolved through the time-dependent dynamics. The Magnus expansion is also often used as a numerical integrator due to its convenient property of guaranteeing state unitarity, even at low orders~\cite{magnusapplications}. However, evaluating the expansion using standard techniques is unfeasible in practice, as they rely on high-order time integrals of commutators of the time-dependent Hamiltonian. The majority of studies that make use of the Magnus expansion limit themselves to at most order four, due to the numerical limitations~\cite{Auer2018,Chakraborty2025gpuaccelerated}.



One of the first studies that used the ME as a numerical integrator was performed in Ref.~\cite{Iserles1999}.  It validates the error scaling of the ME approach by comparing with typical Runge-Kutta integrators, reaching expansion order up to four. More recent works  are able to approximate the evolution under the Magnus expansion using commutator-free methods, achieving at most order eight methods~\cite{Alvermann2011,Auer2018,MorenoCasares2024}. Finally, the authors in \cite{ARNAL2025129563} were able to find a method to efficiently express the continuous Baker-Campbell-Hausdorff formula \cite{dynkin1947,achilles2012early}, a particular instance of the Magnus expansion, in terms of the chained integrals up to order eight. However, they rely on 
intricate analytic integral expressions, the computation of which does not scale favourably for higher orders, and a systematic way to evaluate them is not provided.

The study of Floquet systems revolves around interpreting the dynamics of a time-periodic Hamiltonian as an effective (constant) Hamiltonian  \cite{10.21468/SciPostPhys.2.3.021,Wu2025,Rodriguez-Vega2021Low-frequencyReview,Kalinowski2023Non-AbelianSimulator,WeitenbergTailoringEngineering}. Consequently, the Magnus expansion naturally becomes a fundamental tool in this analysis, being called in this setting the Floquet-Magnus expansion \cite{Milfeld1983-floquetMagnus}.  The study of Floquet dynamics is crucial for the development of quantum technologies as it is often present on the study of quantum materials~\cite{Wilczek2012,LIU2023100705,Eckstein2024}. However, the complexity of the ME limits these studies to only a few orders—-generally up to four. 


In this work, we consider single-control Hamiltonians and solve the high-order time-integrals of commutators involved in the Magnus expansion, reducing the complexity of its calculation to the degrees of freedom of the time-dependent control function. In practice, given a Hamiltonian model, a time-consuming numerical calculation is needed at the start which results in a polynomial expression in the control parameters. This expression efficiently calculates the effective Hamiltonian, evaluating the Magnus expansion up to order twelve within less than a millisecond. This enables fast and accurate simulation of time-dependent quantum systems.

Furthermore, since the expression is analytically differentiable, implementation in optimal quantum control is possible. We propose a method that corresponds to an extension of Gradient Ascent Pulse Engineering (GrAPE) \cite{KHANEJA2005296} for continuous pulses, using the truncated Magnus expansion. Moreover, in this extension, we optimize the pulse duration directly with the fidelity. This work opens the way for new approaches in pulse design for multi-qubit controls using Lie-algebraic techniques.


The article is structured as follows: in \cref{sect:method-background} we introduce the Magnus expansion. As a precursor to the method, we introduce the dynamical Lie algebra and consider the control function to be of polynomial form, corresponding to boxes (a) and (b) of the diagram  in~\cref{fig:graphical-abstract}, respectively. 
This leads to the main result of this work in~\cref{sect:method-polynomial}, displayed in the central box (c) in the diagram.
In \cref{sect:state-intgrator}, we look at the application of our method to quantum simulation (box (d)), study the error scaling with increasing orders, and argue for the high scalability of the method through numerical benchmarking.  In \cref{sect:optimal-control-rydberg},  we showcase an optimal control application for a Rydberg atom-based system, displayed in box (e) of the diagram. Finally, in \cref{sect:conclusion} we conclude and discuss an outlook for future work.

\section{The Magnus expansion} \label{sect:method-background}

Coherent quantum systems evolve according to the time-dependent Schr\"odinger equation~\footnote{in units where $\text{$\hbar=1$}$}
\begin{equation}
    \frac{d}{dt} \ket{\psi(t)} = -i H(t) \ket{\psi(t)},
\end{equation}
 where the Hamiltonian $H(t)$ propagates the state of the system. The unitary state propagator that evolves the state from $t_0$ to $t_f$ is 
 \begin{equation}
    U(t_f, t_0) = \mathcal{T}\exp \left\{ -i \int_{t_0}^{t_f}\hspace{-1mm} ds\, H(s) \right\},
\end{equation}
where $\mathcal{T}$ is the time-ordering operator \cite{wikipedia_ordered_exponential}. In this representation, $U(t_f, t_0)$ is the propagator, encoding the evolution of an initial state $\ket{\psi_0}=\ket{\psi({t_0})}$ into ${\ket{\psi(t_f)} = U(t_f, t_0) \ket{\psi_0}}$. 


The unitary $U(t_f, t_0)$ can be interpreted as the propagator of a constant-in-time evolution of an effective Hamiltonian $H_E(t_f, t_0)$
\begin{equation}
    U(t_f, t_0) = \exp\left\{{-i\, H_E(t_f, t_0)\ (t_f-t_0)}\right\}.
\end{equation} 
The Magnus expansion (ME) provides a constructive approach for calculating
\begin{equation}
    H_E(t_f, t_0)= \frac{1}{t_f-t_0}\lim_{k_M\rightarrow \infty} M^{(k_M)}(t_f, t_0), 
\end{equation}
 as long as the corresponding series converges \cite{magnusapplications}. The series expansion takes the form
\begin{equation}
    M^{(k_M)}(t_f, t_0) := \sum_{k=1}^{k_M} M_k(t_f, t_0),
\end{equation}
with
\begin{align}
    M_1(t_f, t_0) &:= \int_{t_0}^{t_f}\hspace{-3mm} dt_1\ H(t_1),\\
    M_n(t_f, t_0) &:= (-i)^{n-1} \sum_{j=1}^{n-1} \frac{B_j}{j!}\hspace{-1mm} \int_{t_0}^{t_f}\hspace{-3mm} dt_1\, S_n^{(j)}(t_1), \quad n\ge 2,
\end{align}
where $B_j$ is the $j^{th}$ Bernoulli number \cite{bernoulli1713ars}. The $S_n^{(j)}(t)$ operators are defined recursively as
\begin{align}
    &S_{n}^{(1)}(t) :=  [M_{n-1}(t, t_0), H(t)],\\ 
    &S_n^{(n-1)}(t) := \text{ad}_{M_1(t)}^{n-1}(H(t)),\\
    &S_n^{(j)}(t) := \sum_{m=1}^{n-j} [M_m(t, t_0), S_{n-m}^{(j-1)}(t)], \quad 2\le j\le n-1.
\end{align}
with the adjoint operator ${\text{ad}^1_A(B) = [A,B]}$, and $ {\text{ad}_A^n(B)=[A, \text{ad}^{n-1}_A(B)]}$. For illustrative purposes, the first three orders of the Magnus expansion are shown below as time-integrals of commutators of $H(t)$ for varying values of $t$, where $M_k:= M_k(t_f,t_0)$
\begin{align}
    M_1 &= \int_{t_0}^{t_f}\hspace{-3mm}dt_1\, H(t_1) \label{eq:magnus-integral-order1}, \\
    M_2 &= -\frac{i}{2} \int_{t_0}^{t_f}\hspace{-3mm}dt_1 \hspace{-1mm}\int_{t_0}^{t_1}\hspace{-3mm}dt_2\ [H(t_1), H(t_2)],\label{eq:magnus-integral-order2}\\
    M_3 &=- \frac{1}{6} \int_{t_0}^{t_f}\hspace{-3mm}dt_1\hspace{-1mm}\int_{t_0}^{t_1}\hspace{-3mm}dt_2\hspace{-1mm}\int_{t_0}^{t_2}\hspace{-3mm}dt_3 \big([H(t_1), [H(t_2), H(t_3)]]\nonumber\\
    &\hspace{0.38\linewidth}+[H(t_3), [H(t_2), H(t_1)]]\big).\label{eq:magnus-integral-order3}
\end{align}
The Magnus expansion approximates the effective Hamiltonian $H_E(t_f, t_0)$ as long as the series converges. A necessary and sufficient condition for convergence is given by
\begin{equation}
    \int_{t_0}^{t_f} dt\ \| H(t) \|_2 <  \pi,
\end{equation}
where $\|\cdot\|_2$ is the operator spectral norm \cite{Moan2007}. The time $t=t_f-t_0$ where the above condition becomes an equality is called the convergence time $t_*$. Furthermore, the ME is time symmetric, i.e.
\begin{equation}
    M(t_f, t_0) = -M(t_0, t_f),
\end{equation}
which is an important property for numerical integrators, as it guarantees the preservation of geometric quantities, such as state unitarity. It is often considered $t:= t_f$, $t_0 = 0$, thus writing $M(t) := M(t_f,t_0)$. In numerical studies, the series must be truncated at some order $k_{\text{M}}$
\begin{equation}
    M^{(k_M)}(t) := \sum_{k=1}^{k_M} M_k(t).
\end{equation}
Due to the time symmetry of the series, the truncation error at even orders $k_{\text{M}}=2s, s\in\mathbb{N}$ is given by 
\begin{equation}
    \| M^{(2s)} (t)- H_E(t)\| = \mathcal{O}\left(t^ {2s+3}\right). \label{eq:error-scaling-time-symmetry}
\end{equation}
For more details on the Magnus expansion, we refer the reader to the review paper in Ref.~\cite{magnusapplications}. 


\section{Polynomial expression for the Magnus expansion}\label{sect:method-polynomial}
We consider the case of Hamiltonians control models consisting of a time-independent, non-controllable term $A$ and a single controllable term $B$
\begin{equation}
    H(t) = A + d(t) B, \label{eq:hamiltonian-single-control}
\end{equation}
where both $A,B$ are Hermitian operators.
This family of Hamiltonians is often considered in quantum simulation or control problems, with the physical systems ranging from neutral atoms~\cite{Morgado2021QuantumQubits} to superconducting circuits~\cite{Allen2017}. 

 We aim to find the effective Hamiltonian $H_E(t)$ that is generated from the time-dependent dynamics of $H(t)$. The Magnus expansion provides a way to do this for ${t<t_*}$, but it requires the calculation of high-order integrals involving commutators of $H(t)$, similar to the ones seen in \cref{eq:magnus-integral-order1,eq:magnus-integral-order2,eq:magnus-integral-order3}. The direct calculation is not feasible in practice, and so it is necessary to find alternative ways to obtain $H_E$. 

The control function $d(t)$ is represented as a polynomial with maximum degree $m$
\begin{equation}
    d(t) := \sum_{\gamma=0}^m d_\gamma \, \frac{t^\gamma}{\gamma!},
\end{equation}
where $d_\gamma$ are real-valued coefficients of the polynomial $d(t)$ and ${\vec d = (d_0,\ldots, d_m)}$ is its vectorial representation. This finite representation of the control function as a polynomial in time is useful as it allows an analytical expression for the high-order time-integrals involved in the Magnus expansion. As long as the order of the polynomial $m$ is greater than the truncation order of the ME, the dynamics resulting from an arbitrary control function and from its polynomial approximation will be practically equal. This allows us to accurately represent all realistic control functions. This is represented in box (b) of \cref{fig:graphical-abstract}.

The next element of the calculation are the high-order commutators. The effective Hamiltonian generated from the time-dependent dynamics of $H(t)$ belongs to the dynamical Lie algebra which is generated from the operators $A,B$ \cite{gsim}. The dynamical Lie algebra $\mathfrak{g}$ is a vector space that is closed under the Lie bracket ${[A,B]=AB-BA}$, with
\begin{equation}
    \mathfrak{g} := i\ \text{span}_{\mathbb{R}} \langle \{ iA, iB\} \rangle_{\text{Lie}} = \text{span}_{\mathbb{R}} \{L_\mu\}_{\mu=1}^{\text{dim}(\mathfrak{g})}.
\end{equation}
The Lie closure $\langle \cdot \rangle_{\text{Lie}}$  is the set of elements obtained by repeatedly applying the Lie bracket until no new linearly independent elements are obtained \cite{gsim}. The set $\{L_\mu\}_{\mu=1}^{\text{dim}(\mathfrak{g})}$ is a basis of independent Hermitian elements of the Lie algebra $\mathfrak{g}$. We denote $\mathfrak{g}_n$ the dynamical Lie algebra that is generated after applying the Lie bracket at most $n$ times. In the truncated Magnus expansion with order $k_{\text{M}}$, the largest algebra required is $\mathfrak{g}_{k_{\text{M}}}$. Often, the size of this dynamical Lie algebra is substantially smaller than dim$(\mathfrak{g})$, which is advantageous for the efficiency of the algorithm proposed in this work. On the Lie algebra we can define a structure constant $f_{ijk}$ that describes the relation between its elements
\begin{equation}
    [L_i, L_j] = \sum_{k} f_{ijk} L_k,\  L_k \in \mathfrak{g}_n, \ \forall\, L_i, L_j\in \mathfrak{g}_n.
    \label{eq:lie-algebra-structure}
\end{equation}
Thus, when we write the Hamiltonian $H(t)$ in the basis of the elements of this Lie algebra, we know \textit{a priori} all of the commutation relations that are required by the ME. This component of the calculation corresponds to box (a) of~\cref{fig:graphical-abstract}. 

The final step of this process is to find out which integrals of commutators have to be solved for a given expansion order.  We resort to the proposal from Ref.~\cite{Iserles1999} due to its systematic nature. They represent  integrals of commutators of $H(t)$ as binary trees, and the Magnus expansion as a weighted series of binary trees. This approach is explained in detail in~\cref{sect:appx:magnus-tensor-contraction-deduction}. 

Together, these steps yield an explicit expression of the Magnus expansion, where we can quickly obtain the effective Hamiltonian. It is expressed as an element of the dynamical Lie algebra
\begin{equation}
    M^{(k_{\text{M}})}\big(t, \vec d\, \big) = \sum_\mu a^{(k_{\text{M}})}_\mu\big(t, \vec d\, \big) \ L_\mu, \label{eq:lie-sum-ME}
\end{equation}
with truncation order $k_{\text{M}}$,  propagation time $t$, and control functions represented with $\vec d$. The Lie algebra coefficients $a_\mu := a^{(k_{\text{M}})}_\mu\big(t, \vec d\, \big)$ are polynomials in the evolution time $t$ and in the control function coefficients 
\begin{equation} 
    a_\mu  := \sum_{k=1}^{k_{\text{M}}}  \sum_{p=0}^k \hspace{-3.5mm} \sum_{\quad \gamma_1, \ldots,\gamma_p = 0}^m \hspace{-4.5mm} \ \mathbf{S}_{\mu,\vec \gamma} ^{(k,p)}\ t^{k} \prod_{i=1}^p \left(d_{\gamma_i} t^{\gamma_i}\right), \label{eq:coef-lie-ME-0}
\end{equation}
where the index $\gamma_i$ is related to the control coefficient $d_{\gamma_i}$, 
and to the dynamical coefficient $\mathbf{S}_{\mu,\vec \gamma}^{(k,p)}:= \mathbf{S}_{\mu,\gamma_1,\ldots,\gamma_p}^{(k,p)}$, with $ \vec \gamma = (\gamma_1,\ldots,\gamma_p)$. The derivation of this expression is detailed in \cref{sect:appx:magnus-tensor-contraction-deduction}, showing the formula for the coefficients $\mathbf{S}_{\mu,\vec \gamma}^{(k,p)}$. The number of summation terms in this expression can be improved by setting a time truncation order $\Gamma$, that is, neglecting terms in the sum that are smaller than $\mathcal{O}(t^{\Gamma})$. The $\gamma_i$ indices are independent of order, thus we can remove the symmetries in the sum in $\vec \gamma$. This yields an expression with a similar form
\begin{equation}
    a_{\mu} = \sum_{k=1}^{k_{\text{M}}} \sum_{p=0}^k \hspace{-3.5mm} \sum_{ \quad \vec \gamma \in \mathcal{G}_{\Gamma}^{(k,p)}}^m \hspace{-4.5mm} \mathbf{T}_{\mu,\vec \gamma}^{(k,p)} \ t^{k} \prod_{i=1}^p \left(d_{\gamma_i} t^{\gamma_i}\right), \label{eq:coef-lie-ME}
\end{equation}
where the dynamical coefficient $\mathbf{T}_{\mu,\vec \gamma}^{(k,p)}$ is constructed from a sum over the permutations in $\vec \gamma$ of $\mathbf{S}_{\mu,\vec \gamma}^{(k,p)}$, and 
\begin{equation}
\mathcal{G}_{\Gamma}^{(k,p)} := \left\{(\gamma_1, \ldots,\gamma_p)\; \middle|\;\ 
\begin{split}
     &\gamma_i\in \mathbb{N}_0,\\
     & \gamma_i\leq\gamma_{i+1},\\ 
     & k+\sum_{i=1}^p\gamma_i \leq \Gamma
\end{split}\right\}.
\end{equation}

The dynamical coefficients $\mathbf{T}_{\mu,\vec \gamma}^{(k,p)}$ encapsulate the time-dependent dynamics that are generated from $H(t)$, up to a certain order $k_{\text{M}}$. The calculation of these coefficients is quite involved, as explained before, but only needs to be done once for a given control model ($A$ and $B$). 
In fact, a given dynamical coefficient contains contributions from many integrals of commutators that result in the same Lie algebra element $L_\mu$ and which correlate the same monomial in $\vec d$. In essence, this efficiently compresses the complexity of the ME calculation into these dynamical coefficients, instead of the convoluted integrals of commutators previously mentioned in \cref{sect:method-background}. 
The remaining complexity in the calculation is related to the degrees of freedom given by the evolution time $t$ and control coefficients $\vec d$. So, once the dynamical coefficients are obtained, calculating the ME is reduced to a much faster polynomial evaluation.


 

 
With the polynomial form of the ME, it is practical to obtain the effective Hamiltonian for high approximation order. We will show in the following section that this is in fact numerically efficient. The polynomial form of \cref{eq:coef-lie-ME} also allows for analytic derivatives, which we will later showcase its use in optimal control in~\cref{sect:optimal-control-rydberg}.

\section{Quantum Simulation} \label{sect:state-intgrator}

\begin{figure}
    \centering
    \includegraphics[width=1.0\linewidth]{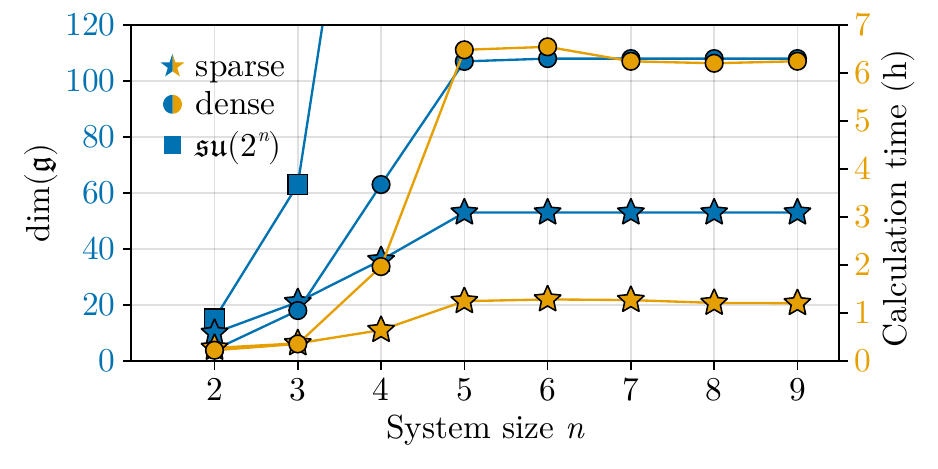}
    \caption{Left axis, blue color: Scaling of the size of the Lie algebra with system size, for the two models and $\mathfrak{su}(n)$, using ${k_{\text{M}}=10}$ and ${\Gamma=12}$. Right axis, yellow color: Scaling of the calculation time, in hours, of the dynamical coefficients as system size increases.}
    \label{fig:calc-time-MEtensor-ryd-scaling}
\end{figure}

\begin{figure}
    \centering
    \includegraphics[width=1.0\linewidth]{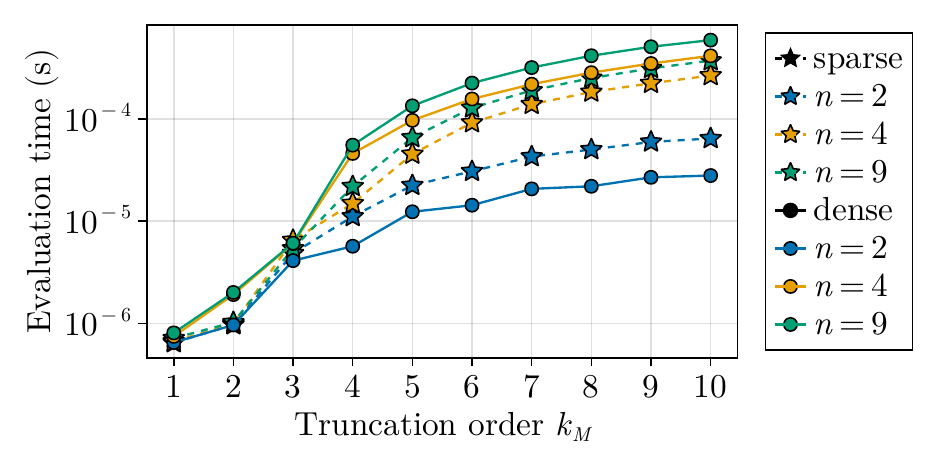}
    \caption{Median evaluation time of the ME polynomial with truncation order $k_M$ and $\Gamma=12$, for both models and various system sizes $n$.}
    \label{fig:contract-time-benchmark}
\end{figure}

For the purpose of quantum simulation we will consider two specific Hamiltonian models: \texttt{sparse} with  nearest-neighbour interactions, and \texttt{dense} with long-range interactions.  
\begin{itemize}
    \item \texttt{Sparse model} - One dimensional transverse field Ising model with $n$ qubits, where the generators of the dynamics are
    \begin{equation}
        A := \sum_{i=1}^{n-1} Z_iZ_{i+1},\quad B:=\sum_{i=1}^n X_i.
    \end{equation}
    \item \texttt{Dense model} - transverse field Ising model with long-range interactions
    \begin{equation}
        A := \sum_{i=1}^{n} \sum_{j>i}^n  J_{ij}Z_iZ_{j},\quad B:=\sum_{i=1}^n X_i,
    \end{equation}
    where $J_{ij}=1/|i-j|$.
\end{itemize}

These models have been selected to allow comparison between the evolution method's performance in the regimes where the dynamics are simple (\texttt{sparse} and/or low $n$) versus when they are complex (\texttt{dense} and/or large $n$). We expect the performance of the method to scale with the size of the dynamical Lie algebra. 

\subsection{Numerical efficiency}\label{sect:numerical-scalability}


Generating the dynamical coefficients $\mathbf{T}_{\mu,\vec \gamma}^{(k,p)}$ and evaluating the polynomials $a_\mu(t, \vec d)$  does not involve states or operators of the full system size. Instead, its numerical complexity scales directly with the complexity of the dynamics of the quantum system, obtained from the Lie algebra, resulting in a scalable method. Moreover, for the generation of the coefficients, we do not require the full Lie algebra, and may only generate it up to the commutation depth $k_\text{M}$ that corresponds to the ME truncation order. Once the Lie algebra and its structure constants are generated, all of the relevant properties of the dynamics of the quantum system are accounted for, within the convergence radius of the Magnus expansion. In \cref{fig:calc-time-MEtensor-ryd-scaling}, we see a strong correlation between the Lie algebra size dim$(\mathfrak{g})$ and the time necessary for generating the dynamical coefficients.
 The size of the dynamical Lie algebra, with limited commutation depth, scales weakly with the number of qubits, and thus generating the dynamical coefficients takes comparable time for different system sizes~\footnote{All of the time-sensitive benchmarking has been done using CPU multithreading, on a system with \efimovspecs.}. 
In \cref{fig:contract-time-benchmark}, we show benchmarks on the evaluation time. Evaluating the ME polynomials is fast, ranging from few $\mu s$ at small orders to $\sim100\,\mu s$ in the larger orders, which is four orders of magnitude faster than the work in Ref.~\cite{ARNAL2025129563}. Moreover, the evaluation time remains nearly constant with increasing system size across all considered models. Access to higher Magnus expansion orders allows for coarser time discretization without compromising accuracy. Consequently, the number of state-propagation steps--and hence the main computational bottleneck--is significantly reduced. This makes the method especially suitable for quantum control applications that demand fast and accurate propagation.



\subsection{Error scaling and time-symmetry}

\begin{figure}
    \centering
    \includegraphics[width=1.0\linewidth]{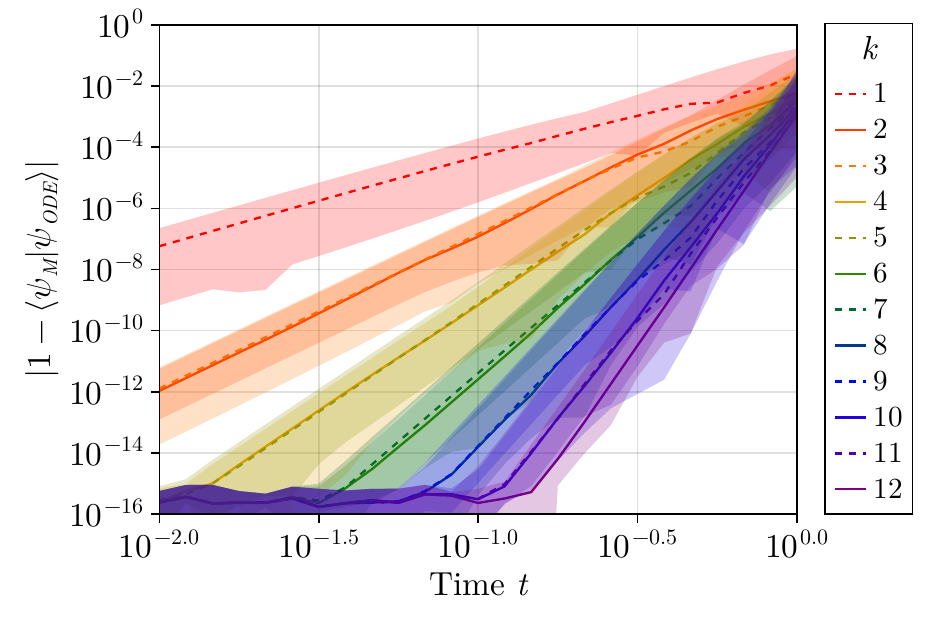}
    \caption{Comparison of state errors at different orders. The control coefficients $d_\gamma$ are sampled from a uniform distribution $d_\gamma\in[-1,1]$.  Used \texttt{sparse} model with $n=3$ and time truncation $\Gamma=14$. Average of 20 samples (random initial state and random controls) is plotted in a line, the minimum and maximum samples are plotted as a band of the respective color. A plateau at around $\sim 2\cdot 10^{-16}$ shows up due to floating-point error.}
    \label{fig:me-ode-error-statistics-sparse-3}
\end{figure}

When truncating the Magnus expansion at even orders ${k_{\text{M}}=2s}$, the error relative to the true effective Hamiltonian is $\mathcal{O}(t^{2s+3})$ (see \cref{eq:error-scaling-time-symmetry}). The ME is used as a propagator in a Hamiltonian simulation setting to obtain the truncation error. This is done by comparing the propagated state when evolving under the effective Hamiltonian 
\begin{equation}
    \ket{\psi_M}:=\exp\left\{-i {M(t,\vec d)}\right\}\ket{\psi_I}
\end{equation}
versus a state propagated with an ODE method $\ket{\psi_{ODE}}$ with sufficient intermediate time-steps~\footnote{We use Verner's 9th order method with 100 time steps, enough to keep the propagation error under floating point accuracy.}, which we consider to be an exact state up to machine precision~\cite{Verner2009,Rackauckas2017,QuantumPropagators}. The initial state is sampled from the uniform Haar distribution \cite{meckes_random_2019}. The results of this comparison for increasing truncation orders are shown in \cref{fig:me-ode-error-statistics-sparse-3}. The log-log plot reveals polynomial scaling, increasing for each even truncation order. 

\begin{figure}
    \centering
     \includegraphics[width=1.0\linewidth]{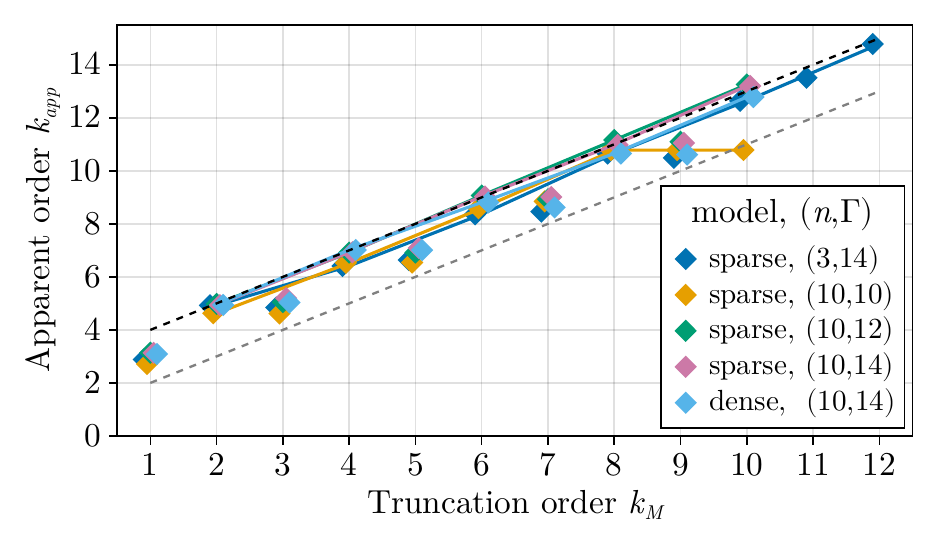}
    \caption{Scaling laws for models \texttt{sparse} and \texttt{dense} with $n=10$, where for the \texttt{sparse} model we also compare the order scaling with $\Gamma$. Dashed lines indicate the time-symmetry threshold at ${k_{app}=k_{\text{M}}+3}$ (black) and the  threshold ${k_{app}=k_{\text{M}}+1}$ (grey). Solid lines connect between markers of the color-corresponding model for even-valued $k_{\text{M}}$. Most models considered follow the time-symmetry scaling $k_{app} \simeq  k_{\text{M}}+3$. } 
    \label{fig:me-ode-error-comparison}
\end{figure}


At each truncation order we fit the error metric to obtain the  corresponding apparent power law $k_{app}$, considering  $\varepsilon_M>10^{-15}$ to suppress floating-point errors. These are shown in~\cref{fig:me-ode-error-comparison} for both models, for different system sizes and $\Gamma$'s. In Ref.~\cite{magnusapplications}, it is shown that $k_{app}=k_{\text{M}}+3$ for even-valued $k_{\text{M}}$ due to the time-symmetry of the expansion. Our numerical tests suggest that this scaling is preserved with our method, as long as $\Gamma\geq k_{\text{M}}+2$.







\section{Optimal control} \label{sect:optimal-control-rydberg}


%

The development of quantum technologies fundamentally relies on the ability to control quantum systems with high precision. This is especially important for quantum computing platforms, where the gate fidelity dictates their usefulness in real-world applications. Contemporary quantum circuits often consist of single- and two-qubit gates, limiting the overall fidelity. It would be preferable if many of these gates could be replaced by an equivalent fast-operating multi-qubit gate. Multi-qubit gate design is especially interesting for neutral atom systems, where multiple atoms simultaneously interact based on their geometric configuration. Quantum optimal control of multi-qubit gates on this platform is a subject of ongoing research \cite{Jandura2022TimeOptimal,PhysRevResearch.5.033052,Evered2023HighFidelity,Fromonteil2024HamiltonJacobiBellman,mohanPraParametrized,Delakouras2025multi,dekeijzer2025consensusbasedqubitconfigurationoptimization}.



However, the design of multi-qubit gates becomes progressively more challenging as the number of qubits increases, due to correspondingly longer pulse durations, a growing number of control parameters, and extended simulation times. The capability to represent the truncated ME as a polynomial function of the control pulse parameters enables numerical optimization of continuous pulses by leveraging analytical differentiation of~\cref{eq:coef-lie-ME}. In our method, the pulse duration is a parameter that is tuned together with the remaining parameters that define the pulse shape. This gives the freedom to include pulse-duration minimization constraints directly as Lagrange multipliers during minimization~\cite{lagrange1788}. Our method corresponds to an extension of GrAPE for continuous pulses. 


On a given quantum control optimization problem, we aim to design a pulse that implements the target gate $G_{tg}$, such that $U(\vec x) \simeq G_{tg}$, where $\vec x$ is the vector of tunable control parameters that define the pulse and $U(\vec x)$ is the unitary propagator of the pulse. The optimization of the state trajectory consists of determining $\vec x$ that minimizes a cost function $J[\vec x]$. In this work we use 
\begin{equation}
    J[\vec x] := \frac{1}{2} \sum_{r=1}^R \left(1-\text{Re}\langle \phi_r |U(\vec x)| \psi_r \rangle \right),
\end{equation}
where $|\psi_r\rangle$ and $|\phi_r\rangle$ are the initial state and target state respectively, for the trajectory indexed by $r$, and $R$ is the number of trajectories that are being optimized. 
A suitable set of initial states $|\psi_r\rangle$ must be chosen for the design of a target gate $G_{tg}$. This set may be a complete basis of the Hilbert space, but often it is possible to use a much smaller set while still addressing all of the symmetries present in the dynamics. Once these are established, the set of target states is uniquely determined $|\phi_r\rangle=G_{tg}|\psi_r\rangle$.

We choose the problem of designing parametrized multi-qubit gates for neutral atoms to showcase the advantages of the method. Namely, we aim to show that we can take advantage of the time-adaptive property of the method to efficiently design gates, by using previous pulse-shapes as initial guesses for gates with similar parameters.

\subsection{Hermite spline control pulse}\label{sect:hermite-spline}

The convergence time of the Magnus expansion is often below the time necessary to design a useful gate. Therefore, we concatenate multiple ME propagation segments to form a longer pulse. Each ME segment is comprised of a polynomial-shape control function, in contrast with GrAPE, where the propagation consists of a sequence of constant step-functions. 
Several options exist for selecting both the representation of the control pulse and its parameters. In this work, we represent and optimize the control pulse as an Hermite spline \cite{fageot_support_2019}. This representation assures continuity for the whole control function, and can be easily converted into the underlying polynomial representation with which we have described the ME. 

\begin{figure}
    \centering
    \includegraphics[width=0.8\linewidth]{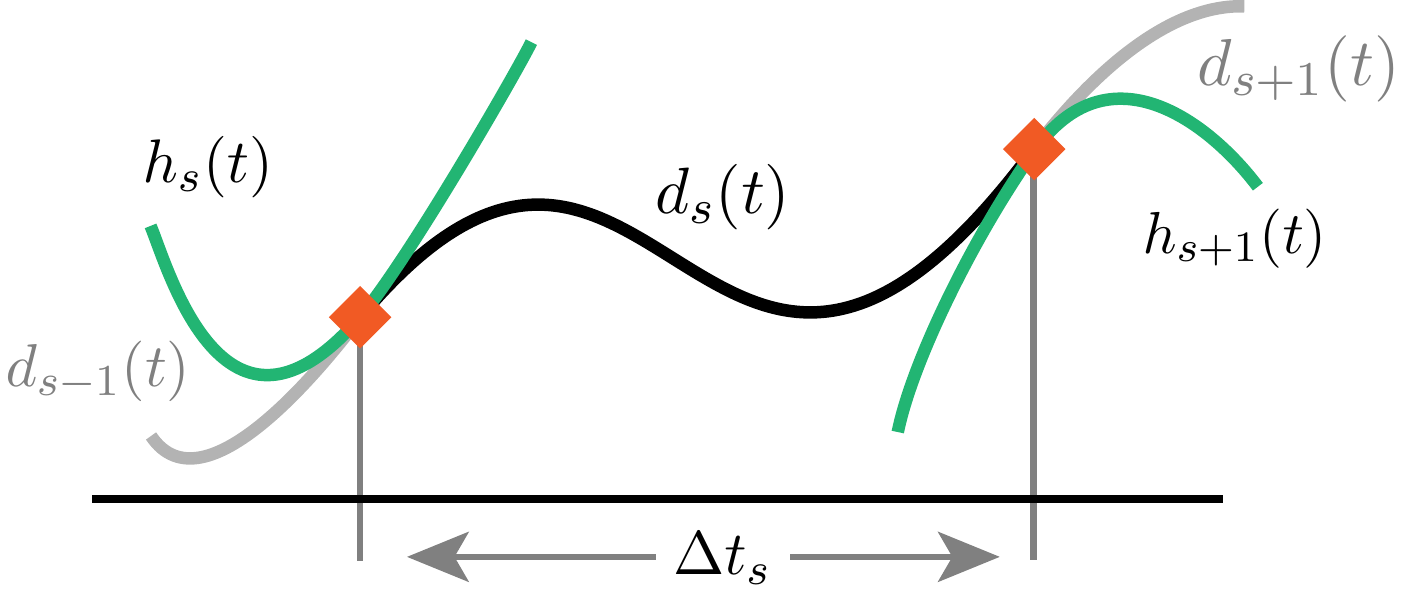}
    \caption{Segment $s$ of an Hermite spline. The Hermite nodes are shown as orange diamond markers, and delimit a time-interval with duration $\Delta t_s$. The Hermite polynomials $h_s(t)$, $h_{s+1}(t)$ at these nodes are represented as green lines. The line in black which connects the orange markers is the polynomial that defines the shape of the spline on this interval. }
    \label{fig:hermite-segments}
\end{figure}

An Hermite spline is a function with class $C^L$ that consists of a sequence of segments that split the control function into $S$ time intervals. The segment at time interval $s$ has a duration of $\Delta t_s$. At the start and end of the segment $s$ there exist Hermite nodes that define a polynomial of degree $L$. These polynomials are denoted $h_s(t)$ and $h_{s+1}(t)$, respectively. The construction of an Hermite spline involves determining the polynomial $d_s(t)$ whose $l$-order derivatives match the ones at the start and end of the segment  
\begin{align}\label{eq:transform-hermite-poly-1}
    &\left. \frac{d^l}{dt^l} d_s(t) \right|_{t=0} = \left. \frac{d^l}{dt^l} h_s(t) \right|_{t=0},\\
    &\left. \frac{d^l}{dt^l} d_s(t) \right|_{t=\Delta t_s} = \left. \frac{d^l}{dt^l} h_{s+1}(t) \right|_{t=0},\label{eq:transform-hermite-poly-2}
\end{align}
with $l=0,\ldots,L$. The polynomial function $d_s(t)$ is of polynomial degree $2L+1$, since the number of constraints on the segment is $2L+2$, and is fully determined with the duration $\Delta t_s$ and the polynomial coefficients $\vec c_s=(\Delta t_s, d_0^{(s)}, \ldots, d_L^{(s)})$. An illustration of one of these segments is shown in \cref{fig:hermite-segments}. A Hermite spline with class $C^L$ and $S$ segments is fully determined with $S+1$ nodes $h_s(t)$ and $S$ time-intervals $\Delta t_s$. Each node has $L+1$ parameters. Thus, the number of parameters for such a spline is $(S+1)\, (L+1) + S$. These are the Hermite spline parameters $\vec h$, in contrast with the segment polynomial coefficients $\vec c=(\vec c_1, \ldots, \vec c_S)$ that define the polynomial function $d_s(t)$ for each of the segments $s=1,\ldots,S$.

The Hermite spline parameters $\vec h$ are the controls that are used on the numerical optimizer. Since the ME is expressed in terms of the segment polynomial coefficients $\vec c$, the Hermite spline parameters $\vec h$ are transformed into the segment polynomial coefficients $\vec c$ through \cref{eq:transform-hermite-poly-1,eq:transform-hermite-poly-2}. The Jacobian of this transformation is required for gradient-based optimization. The mathematical details of this transformation are discussed in \cref{appx:sect:optimal-control}.

\subsection{Rydberg atoms}

The design problem of the multi-controlled phase gate for Rydberg atoms has been tackled in multiple works~\cite{Jandura2022TimeOptimal,Evered2023HighFidelity,Fromonteil2024HamiltonJacobiBellman,mohanPraParametrized,Delakouras2025multi}. Here, we focus on the parametrized variant
\begin{equation}
    G_{tg}=C_kP(\phi) := (\mathbb{I} - (|1\rangle\langle1|)^{\otimes n}) e^{i\phi} + (|1\rangle\langle1|)^{\otimes n}, \label{eq:ckphase-gate}
\end{equation}
where $n=k+1$ is the number of atoms. The typical Hamiltonian considered when studying Rydberg atoms is
\begin{equation}
    H(t) = \frac{\Omega(t)}{2}\sum_{i=1}^n  X_i -  \Delta(t) \sum_{i=1}^n|r\rangle_i\langle r|_i + \sum_{i>j}V_{ij} |rr \rangle_{ij}\langle rr |_{ij},
\end{equation}
with 
\begin{align}
    X_i &:= |0\rangle_i\langle 0|_i + |r\rangle_i\langle 1|_i + |1\rangle_i\langle r|_i,
\end{align}
and where $\Omega(t)$ is the Rabi frequency of a global rotation, $\Delta(t)$ is the corresponding off-resonance detuning of the transition, and $V_{ij}$ is the Van der Waals inter-atomic potential, thus scales with $1/R_{ij}^6$, $R_{ij}$ is the distance between atoms $i,j$ \cite{Morgado2021QuantumQubits}. 


When designing gates for this platform it is common to consider the regime where $V_{ij}\gg \Omega$, called the blockade regime. In the perfect blockade scenario there is a symmetric, all-to-all blockade interaction, which we can represent with a simpler Hamiltonian control model \cite{Jandura2022TimeOptimal} \begin{equation}
    H(t)/\Omega = \frac{1}{2}\sum_{i=1}^n X_i Q_{{[i]}} \ +\ d(t) \frac{1}{2} \sum_{i=1}^n Z_i,  \label{eq:hamiltonian-simple-rydberg}
\end{equation}
where 
\begin{align}
    Q_{[i]}&:=\prod_{\substack{j=1\\j\ne i}}^nQ_j,\quad Q_j :=  |0\rangle_j\langle 0|_j + |1\rangle_j\langle 1|_j,\\
    Z_i &:= |0\rangle_i\langle0|_i + |1\rangle_i\langle1|_i -|r\rangle_i\langle r|_i.
\end{align}
The Rabi frequency is kept constant $\Omega(t)=\Omega$, and the detuning control function $\Delta(t)$ is rescaled accordingly $d(t) = \Delta(t)/\Omega$. The control function $d(t)$ takes the form of an Hermite spline, which corresponds to the control pulse that is to be optimized. We focus only on a subset of the Hilbert space of states
\begin{equation}
    |\psi_i \rangle := |1 \rangle^{\otimes i} |0\rangle^{\otimes (n-i)}, \quad i =0,\ldots, n,
\end{equation}
as the system Hamiltonian remains invariant under atom permutations, faithfully representing the full Hilbert space. 

In gate design for Rydberg atoms it is common to consider the availability of single qubit gates that can act before and/or after an entanglement-generating interaction \cite{Fromonteil2024HamiltonJacobiBellman,maskara2023programmablesimulationsmoleculesmaterials}.  Then, the parametrized gate sequence
\begin{equation}
    U(\vec h, \theta) := \prod_{i=1}^n R_{Z_i}(\theta)  \prod_{s=1}^S D_s(\vec c_s(\vec h_s))
\end{equation}
is used for the design of the target gate,
where $\vec h$ are the Hermite spline parameters as described in~\cref{sect:hermite-spline}, and $\theta$ is the rotation angle for the single-qubit rotations
\begin{equation} 
    R_{Z_i}(\theta) = \mathbb{I}\cos ({\theta}/{2})\  -i (\mathbb{I} - 2|{1}\rangle_i\langle{1}|_i) \sin(\theta/2)
\end{equation} that act on atom $i$, and 
\begin{equation}
    D_s(\vec c_s) := \exp \left\{ -i M^{(k_M)}(\vec c_s) \right\}
\end{equation}
is the unitary propagator obtained from the truncated Magnus expansion from~\cref{eq:lie-sum-ME}. 

The numerical optimization requires the derivatives of the propagator ${\nabla_{\vec h} U(\vec h, \theta)}$ and ${\partial_\theta  U(\vec h, \theta)}$. The method to obtain of these is shown in \cref{appx:sect:optimal-control}.

\subsection{Numerical Optimization}

\begin{figure}
    \centering
    \includegraphics[width=0.99\linewidth]{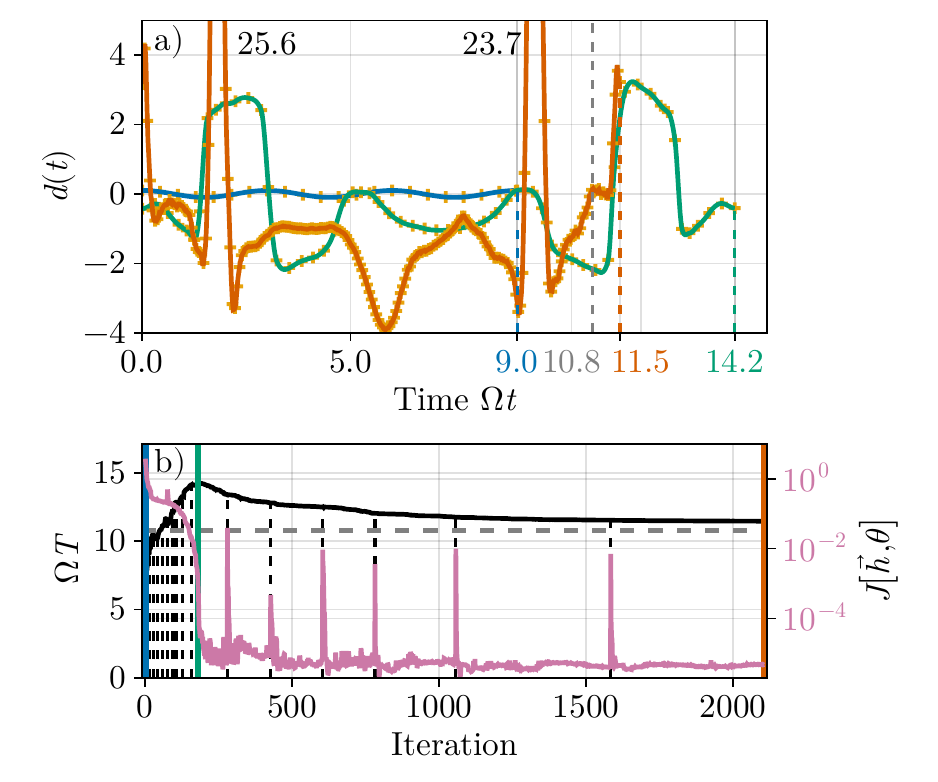}
    \caption{Optimization of the $C_2Z$ gate, with $(T_{min}, T_{max}, \lambda_T, \epsilon_M)=(4.0, 18.0, 10^{-3}, 10^{-6})$. (a) Three pulses are shown above for iterations $0, 180$ and $2106$, with respective colors blue, green and orange. The orange pulse has two marked high value peaks which fall outside of the plotting region. The '$+$' markers in yellow show where the Hermite nodes are placed for that spline. (b) A black line shows how the dimensionless pulse duration $\Omega T$ progresses along the optimization. Vertical dashed lines in black indicate where the threshold on \cref{eq:divergence-threshold} is triggered. The cost function $J[\vec h, \theta]$ is plotted in pink. } 
    \label{fig:c2z-optimization-example}
\end{figure}

For numerical optimization of the cost function $J[\vec h, \theta]$ we have used BFGS with box constraints on the time intervals \cite{mogensen2020optim}. The pulse duration is constrained to $T\in [T_{min}, T_{max}]$ by imposing constraints in the time intervals of the Hermite spline $T_{min}/S \leq \Delta t_s\leq T_{max}/S$. During the optimization the truncation error of the ME is estimated with
\begin{equation}
    \varepsilon_{M}^2 := {\sum_\mu \left(a_\mu^{(k_M)}\right)^2 \|L_\mu\|^2}  \ge \|M_{k_M}\|^2,
\end{equation}
where $\|L_\mu\|$ is the $\ell_1$ operator norm of the Lie algebra element $L_\mu$, computed once before optimization. We set a threshold $\varepsilon_*$ where, whenever
\begin{equation}
    \sum_{s=1}^S\varepsilon_M^{(s)} >\varepsilon_*, \label{eq:divergence-threshold}
\end{equation}
the optimization is stopped in order to increase the number of nodes in the spline, until $\varepsilon_M \leq 0.1 \, \varepsilon_*$, with $\varepsilon_M^{(s)}$ the truncation error at segment $s$. This ensures the truncation error remains bounded while the spline stays nearly unchanged. The downside is that optimization has to be interrupted as the number of Hermite nodes increases, changing $\vec h$. The optimization converges successfully once $J \leq \varepsilon_M$. 
The duration of the pulse is improved by including a Lagrange multiplier term
\begin{equation}
    J_{T} := \lambda_T \sum_{s=1}^{S} \Delta t_s = \lambda_T\, T
\end{equation}
to the cost function, with $\lambda_T$ tuned heuristically.

\begin{table}[]
    \centering
    \begin{tabular}{c|c c c c}
        Iteration & $\quad \Omega T\quad$ & $S$ & dim($\vec h$)  & $J[\vec h, \theta]$  \\\hline
        0 & 9.00 & 21 & 65 & 3.74\\
        180 & 14.2 & 58 & 176 & $9.84 \times 10^{-4}$\\
        500 & 12.6 & 95 & 287 & $5.05 \times 10^{-6}$ \\
        1000 & 11.8 & 123 & 371 & $5.64 \times 10^{-6}$\\
        2106 & 11.5 & 208 & 626 & $4.91 \times 10^{-6}$\\ \hline
    \end{tabular}
    \caption{Parameters and properties of several control pulses for realizing the $C_2Z$ gate.}
    \label{tab:pulses-properties}
\end{table}

In \cref{fig:c2z-optimization-example},  an example is shown of how the control pulse changes throughout the optimization, with the $C_2Z$ gate as the target. The control pulse is an Hermite spline with Hermite order $L=1$. The bounds $\Omega T\in [4.0, 18.0]$ are applied in this optimization. We have observed that with lower $T_{max}$ bounds, the optimizer requires more iterations to find an adequate control pulse, and often with $J>10^{-3}$. The optimization starts with $d(t)=0.1\cos(2\pi t/3)$, with the duration $\Omega T=9.0$. As the optimization progresses, the total truncation error of the ME surpasses the threshold, leading to a resampling of the spline of the control pulse, increasing the number of nodes and the number of control parameters. At iteration $180$, the cost function $J$ decreases substantially and remains at $\sim 10^{-6}$ for the remainder of the optimization. This plateau is likely due to errors in the gradient of the propagator. From there onwards, the pulse duration is gradually improved through the Lagrange multiplier term $J_T$, leading to an increase of the sharpness of the peaks of the control pulse. The final pulse duration is $\Omega T=11.5$, which is $6\%$ slower than the time-optimal pulse with $\Omega T=10.8$~\cite{Evered2023HighFidelity}. This difference likely arises because we optimize the laser detuning $\Delta (t)$ instead of the phase~\footnote{The comparison has its limitations. We are solving for a different optimization problem. An intuition for this distinction comes from the fact that the derivative of the phase of the laser is its detuning, and as such any discontinuity of the phase leads to a Dirac-delta function.}.  Explicit values related to the marked pulses for their duration $\Omega T$,  number of segments $S$, the number of pulse parameters dim($\vec h$) and cost function $J[\vec h, \theta]$ are shown in~\cref{tab:pulses-properties}. As the optimization progresses, and the control pulse becomes more detailed, the number of parameters increases accordingly. It is comparable with the $400$ of discrete time-steps used in the time-optimality study in Ref.~\cite{Jandura2022TimeOptimal}.

We direct our attention now to the optimization of control pulses for the $C_4P(\phi)$ gate family. This is inherently a more challenging problem due to the increased number of qubits. Prior work on time-optimal pulses has, to the best of our knowledge, only considered a single member of this family: the $C_4Z = C_4P(\pi)$ gate~\cite{Evered2023HighFidelity}.
Beyond the increased computational cost of simulation and gradient evaluation, the control pulses themselves exhibit greater structural complexity, requiring more spline segments to represent accurately. This results in a higher-dimensional optimization landscape. 

\begin{figure}
    \centering
    \includegraphics[width=0.99\linewidth]{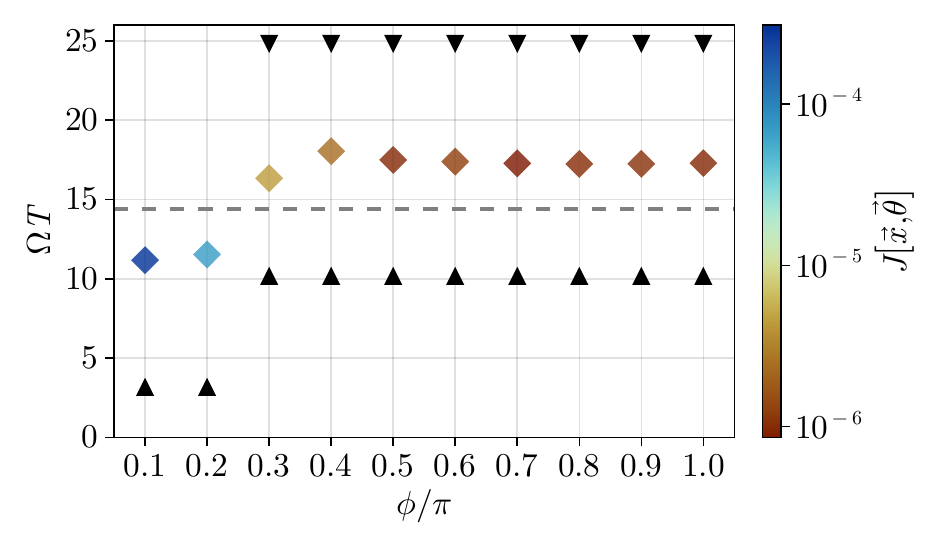}
    \caption{Pulse durations and costs for the different angles. The colors of the markers indicate the respective costs, related to the color bar on the right. The size of the markers at a given $\phi/\pi$ showcase the progress of the pulse duration and cost along the iterations. Smaller markers indicate earlier iterations. The grey line indicates the duration of the time-optimal pulse at $\phi=\pi$ from Ref.~\cite{Evered2023HighFidelity}. For $\phi<\pi$, we expect lower optimal times. The upper and lower triangle markers indicate the imposed pulse duration bounds $T_{max}, T_{min}$, respectively. }
    \label{fig:optim-ckp-traj}
\end{figure}

Each round of optimization proceeds similarly to the $C_2Z$ case. Our strategy is to optimize the gate family in increments of $\Delta\phi = \pi/10$, using warm-starting: the optimized pulse for $C_4P(n\Delta\phi)$ serves as the initial guess for $C_4P((n+1)\Delta\phi)$.  We begin by finding a pulse for the $C_4P(\Delta\phi)$ gate with $\varepsilon_* = 10^{-4}$, tuning $T_{\min}$ and $T_{\max}$ heuristically. At this stage we set $\lambda_T = 0$, as pulse duration is not yet a concern. Once $J \leq \varepsilon_*$, we refine by tightening to $\varepsilon_* = 5 \cdot 10^{-6}$ and enabling duration optimization with $\lambda_T = 10^{-4}$. This refinement procedure is then repeated for each subsequent gate up to $C_4P(\pi) = C_4Z$. For $\phi/\pi \geq 0.3$, we update the duration constraints to improve optimizer convergence. The optimization results are shown in \cref{fig:optim-ckp-traj}. The $C_4Z$ pulse obtained is approximately $20\%$ longer than the time-optimal result of Ref.~\cite{Evered2023HighFidelity}, and this difference we attribute again to optimizing the laser detuning instead of its phase. The remaining gates in the interval $\phi/\pi \in [0.3, 1.0]$ exhibit similar durations, a trend also observed for two- and three-qubit controlled-phase gates~\cite{Jandura2022TimeOptimal,mohanPraParametrized}.

\section{Discussion and Outlook}\label{sect:conclusion}

In this work, we express the solution operator from the Magnus expansion in polynomial form for Hamiltonians consisting of a time-independent drift and a single control term.
This is achieved by systematically solving the high-order integrals of commutators required by the Magnus expansion. 
This polynomial form condenses the dynamics into a set of coefficients, shifting the computational complexity of the Magnus expansion to the control degrees of freedom alone.
This method enables numerical calculation of the Magnus expansion in hundreds of microseconds, up to twelfth order, while being four orders of magnitude faster than previous approaches.
Naturally, this application is well suited for tasks involving simulation of time-dependent quantum systems in classical computers, but can also be an important analytical tool for the study of these systems. 

We further demonstrate the method's applicability to quantum optimal control by designing multi-qubit gates for Rydberg-atom systems. The approach enables optimization of continuous control pulses represented as splines while maintaining a similar number of parameters when compared with the state-of-the-art. As the pulse duration is optimized jointly with the pulse shape, suitable solutions require less fine-tuning compared to standard GrAPE-based techniques. Using this framework, we successfully design pulses for the five-qubit phase gate family, leveraging the adaptive-time capabilities of our method.

\section{Code availability}

The Julia code to generate the dynamical coefficients, evaluate the respective Magnus Expansion polynomial and its derivatives can be found in Ref.~\cite{magnus-polynomial-rualito}. It also contains the code for the optimal quantum control methods adopted in this work. 

\section{Acknowledgements}

We thank Robert de Keijzer, Jasper Postema, Jasper van de Kraats, Madhav Mohan, Pim Kersbergen  and Emre Akaturk for fruitful discussions. This research was financially supported by 
the Dutch Ministry of Economic Affairs and Climate Policy (EZK), as part of the Quantum Delta NL program, 
and the Horizon Europe program HORIZON-CL4-2021-DIGITAL-EMERGING-01-30 via Project No. 101070144 (EuRyQa).

\medskip

\bibliographystyle{abbrv}
\bibliographystyle{unsrt}
\bibliography{library}

\newpage


\appendix
\onecolumngrid

\section{Representing the Magnus expansion as a polynomial}\label{sect:appx:magnus-tensor-contraction-deduction}

The Magnus expansion can be expressed as a sum of integrals of commutators represented as binary trees, proposed by Ref.~\cite{Iserles1999}. The $k^{th}$ order term of the ME is
\begin{align}
    M_k(t) &= (-i)^{k-1}  \sum_{\tau \in \mathbb{T}_{k-1}} \alpha(\tau) \int_0^t dt_1 \cdots \int_0^{t_{\tau}} dt_k [H(t_1) \cdots H(t_k)]_\tau, \label{eq:appx:magnus-expansion-term-tree-repr}
\end{align}
where  $\mathbb{T}_k$ is the set of binary trees representing integrals and commutators of order $k$, $t_{\tau}\in\{ t, t_1, \ldots,t_{k-1} \}$ depending on $\tau$, $[H(t_1) \cdots H(t_k)]_\tau$ represents chained commutators in the structure of the tree $\tau$, and $\alpha(\tau)$ is a weight given to the tree $\tau$. The upper-times in each of the integrals in the chain depend on the tree $\tau$. The specifics of how to construct such chain of integrals and commutators is given below. 

\subsection*{Binary trees as integrals of commutators}\label{sect:bin-trees}

The set of binary trees is generated recursively according to 
\begin{equation}
    \mathbb{T}_k := \Bigg\{ \tafix \,:\, \tau_1 \in \mathbb{T}_{m_1}, \,\tau_2 \in \mathbb{T}_{m_2},\, m_1+m_2 = k-1 \Bigg\},
\end{equation}
where the trivial case is $\mathbb{T}_0 = \{\tikz  \circbfix ;\ \}$. Then the first few sets of trees are
\begin{align}
    \mathbb{T}_1 &= \Bigg\{ \tbfix \Bigg\},\\
    \mathbb{T}_2 &= \Bigg\{ \tcfix, \tdfix \Bigg\},\\
    \mathbb{T}_3 &= \Bigg\{ {\tifix}, {\tgfix}, {\tefix}, {\tffix}, {\tjfix} \Bigg\}.
\end{align}
We can identify any tree $\tau$ in a unique way 
\begin{equation*}
    \tau = {\tuniquefix}
\end{equation*}
where each $\tau_l$ are the subtrees of $\tau$. For each of the trees in $\mathbb{T}_k$ there is a corresponding weight $\alpha(\tau)$ on the ME given by
\begin{equation}
    \alpha(\tau) = \frac{B_s}{s!} \prod_{l=1}^s \alpha(\tau_l),
\end{equation}
where the trivial tree has weight $\alpha(\tikz  \circbfix ;)=1$ and $B_s$ is the $s^{th}$ Bernoulli number \cite{bernoulli1713ars}. Each tree represents the integral and commutation of the Hamiltonian evaluated at different times. This is done recursively according to
\begin{equation}
    H_\tau(t) = \left[ \int_0^tds\ H_{\tau_1}(s),\ H_{\tau_2}(t) \right]\quad \text{with} \quad \tau = \tafix. \label{eq:appx:recursive-tree-integral}
\end{equation}
The symbolic tree integral from \cref{eq:appx:magnus-expansion-term-tree-repr} then corresponds to the integration of \cref{eq:appx:recursive-tree-integral}
\begin{equation}
    \int_0^t dt_1\ H_{\tau}(t_1) = \int_0^t dt_1 \cdots \int_0^{t_{k_\tau}} dt_k [H(t_1) \cdots H(t_k)]_\tau.
\end{equation}
For illustrative purposes, we follow with a few examples in sequence. The single tree in $\mathbb{T}_1$ represents the integral
\begin{equation}
    \tbifix\ \sim\ \int_0^t dt_1 H_{\scalebox{0.5}{\tbfix}}(t_1) = \int_0^t dt_1 \left[\int_0^{t_1} dt_2 H(t_2), H(t_1)\right] =   \int_0^t dt_1 \int_0^{t_1} dt_2 \left[ H(t_2), H(t_1) \right].
\end{equation}
which is the second term of ME. Then, following to the second element of $\mathbb{T}_2$
\begin{equation}
    \tdifix \ \sim\ \int_0^tdt_1 H_{\scalebox{0.5}{\tdfix}}(t_1) =\int_0^{t}dt_1 \left[ \int_0^{t_1} dt_2 H(t_2), H_{\scalebox{0.5}{\tbfix}}(t_1) \right]  = \int_{0}^t dt_1 \int_0^{t_1} dt_2 \int_0^{t_1} dt_3 [ H(t_2), [H(t_3), H(t_1)] ].
\end{equation}
Another more complex case that shows up at fourth order is illustrated here
\begin{equation}
    \teifix \quad\sim\quad \int_0^t dt_1 \int_0^{t_1} dt_2 \int_0^{t_2}dt_3 \int_0^{t_2} dt_4 \left[ \left[  H(t_3),\left[H(t_4), H(t_2) \right]\right], H(t_1) \right]. \label{eq:tree-example-4}
\end{equation}
Each vertical line represents time integration, and each bifurcation corresponds to a commutation. The commutators at the root of the tree are computed from the leaves downward (shown as $\tikz  \circbfix\  ;$). The tree also determines the integration limits. Each leaf on the tree is uniquely associated with an integration time $t_i$. 

\subsection*{Separating the integrals from the commutators} \label{sect:method}


The high order terms of ME involve an increasingly complex series of integrals and commutators. Evaluating $H(t)$ independently on these integrals becomes numerically unfeasible if no special treatment is taken. We have a single-control Hamiltonian
\begin{equation}
    H(t) = A + d(t) B,
\end{equation}
which allows us solve the high-order commutators independently of the integrals. The core of the method involves generating a basis for the Lie algebra generated by $\{A, B\}$, plus approximations neglecting higher powers in the time $t$. To help illustrate the process, consider the third order commutator that is related to the second tree in $\mathbb{T}_2$
\begin{equation}
\begin{split}
    [H(t_1) \cdots H(t_3)]_{\scalebox{0.5}{\tdfix}} &=\left[ H(t_1), [H(t_2), H(t_3)] \right] \\
    &= [A,[A, d(t_3)B]] \\
    &+ [A,[d(t_2)B, A]]\\
    &+[d(t_1)B, [A, d(t_3)B]]\\
    &+ [d(t_1)B, [d(t_2)B,A ]].
\end{split}
\end{equation}
In this calculation we distribute the sums for $A,B$ that are inside of the chained commutator. In each entry of the commutator one can place either $A$ or $B$, and through distribution they are placed in every combination. For a $k$-order chained commutator, there are $2^k$ options for placement of $A,B$, although in some of them, the commutation result is zero due to self-commutation ($[A,A] = [B,B] = 0$). Depending on which entry the operator $B$ is placed, the control $d(t_\circ)$ is evaluated at a different time index, as illustrated above. Then, we adopt the notation
\begin{equation}
    [(A), ( d(t_\circ)B)]_{\tau,i}
\end{equation}
for the element $i$ of a distributed sum, with the commutator structure given from the tree $\tau$. Here are examples of the representation of the terms shown in the calculation above, with $i=i_k\cdots i_1$ represented in base-2
\begin{align}
    [(A), ( d(t_\circ)B)]_{\tau, 001_{2}} &= [A,[A, d(t_3)B]],\\
    [(A), ( d(t_\circ)B)]_{\tau, 010_{2}} &= [A,[d(t_2)B, A]],\\
    [(A), ( d(t_\circ)B)]_{\tau, 101_{2}} &=[d(t_1)B, [A, d(t_3)B]],\\
    [(A), ( d(t_\circ)B)]_{\tau, 110_{2}}&=[d(t_1)B, [d(t_2)B,A ]].
\end{align}
Then, a generic commutator chain that comes from a tree $\tau$ is given by
\begin{equation}
    [(A), ( d(t_\circ)B)]_{\tau,i} = \sum_{i_k=0}^1 \cdots \sum_{i_1=0}^1  [(A), (d(t_\circ) B)]_{\tau, i_k\cdots i_1}= \sum_{i=0}^{2^k-1} [(A), (d(t_\circ) B)]_{\tau, i}
\end{equation}
where some of these terms evaluate to zero due to self-commutation. The algorithm to evaluate each term of this sum is explained in \cref{fig:binary-tree-commutators}. The Lie algebra is expressible only in relation to the commutators of $A,B$, and so we need to separate the time dependence that originates from $d(t_{\circ})$. Due to the linearity of the commutator, the chain can be written as the following sum
\begin{equation}
    [H(t_1) \cdots H(t_k)]_\tau = \sum_{i=0}^{2^k-1}\left( \prod_{\{j \in [k] : i_j=1\}} d(t_j) \right) [(A), (B)]_{\tau, i} \label{eq:appx:commutator-chain-separation}
\end{equation}
where the product over the set $\{j \in [k] : i_j=1\}$ can be read as \textit{all $j\in\{1, \ldots,k\}$ such that the $j^{th}$ digit of $i$ in base 2 is 1}. The commutator chain $[(A), (B)]_{\tau, i}$ can then be efficiently determined by the Lie algebra structure tensor, which we will describe below. The time-dependent term that is the product of $d(t_\circ)$ will be integrated under the chained integral defined from the tree $\tau$.


\subsection*{Generating Lie Algebras} \label{sect:lie-algebra}

The chain of commutators $[(A), (B)]_{\tau, i}$ can be efficiently computed a priori by generating an orthogonal basis of operators via the Gram-Schmidt process. The Lie algebra set $\mathfrak{g}_k$ of order $k$ contains all of the independent operators that are generated from the generator set $G=\{A, B\}$ up to commutation order $k$. The first order set $\mathfrak{g}_1=\{L_1, L_2\}$ where
\begin{equation}
    L_1 = A, \qquad L_2 = B - (\text{proj}_{ L_1 } B)
\end{equation}
where $\text{proj}_{ L_1 } B$ is the projector of operator $B$ into $L_1$. When $B, L_1$ are matrices, the projection is computed on the vectorized representations of $B, L_1$. The set $\mathfrak{g}_k$ is generated recursively as follows

\begin{equation}
    \mathfrak{g}_{k+1} = \left\{ L_a : L_a = [g, L_b] - \text{proj}_{\{L_1, \dots, L_{a-1}\}} [g, L_b],\ L_b \in \mathfrak{g}_k,\ g\in G \right\} \backslash \{0\} \label{eq:independent-lie-algebra}
\end{equation}
where $\text{proj}_{\{L_1, \dots, L_{a-1}\}} [g, L_b]$ is the projection of $[g, L_b]$ into the orthogonal space generated by  $\{L_1, \dots, L_{a-1}\}$. The full Lie algebra generated by $G$ is then the union of of all of the sets of each order $\mathfrak{g} = \cup^{M}_{m=1} \,\mathfrak{g}_m$, where the recursive depth is at most $M$, where $\mathfrak{g}_M=\emptyset$. We say that if a term $Q \in \mathfrak{g}_k$, then $Q$ shows up at commutation depth $k$ on the Lie algebra. A cut-off is implemented when calculating this numerically, neglecting terms $L_a$ where $\|L_a\|_1 < \varepsilon_{L} \min(\|A\|_1, \|B\|_1)$, with $\varepsilon_{L}=10^{-5}$. For a finite number of qubits, the number of elements in $\mathfrak{g}$ is also finite, and may correspond to the full Hilbert space of operators. 

The structure constant $f_{ijk}$ of the Lie algebra describes all of the necessary relations between its elements
\begin{equation}
    [L_i, L_j] = \sum_{L_k \in\mathfrak{g}} f_{ijk}\ L_k, \ \forall \ L_i,L_j \in \mathfrak{g} \label{eq:structure-tensor}
\end{equation}
which is enough to compute all of the commutation relations that show up in ME. The calculation of $f_{ijk}$ takes $\mathcal{O}(\text{dim}(\mathfrak{g})^2)$ commutator evaluations. Each of those is projected back to the Lie algebra $\mathfrak{g}$ using the Gram-Schmidt process. Any chain of commutators can be represented as an element of the Lie algebra 
\begin{equation}
    [(A), (B)]_{\tau, i} = \sum_{L_\mu \in \mathfrak{g}} \beta_{\mu,\, i \tau} L_\mu. \label{eq:tree-commutators-lie-algebra}
\end{equation}
See \cref{sect:appx-lie-algebra-over-tree} for an illustration of this calculation.


\subsection*{Polynomial representation} \label{sect:solving-trees}

Now with the Lie algebra relations , the integral over the polynomial function needs to be determined. 
Rewriting the sequence of integrals defined over the tree $\tau\in\mathbb{T}_{k-1}$, for convenience
\begin{equation}
    \fint_0^t dt_{[\tau]}  :=\int_0^t dt_1 \cdots \int_0^{t_\tau} dt_k.
\end{equation}
The equation \cref{eq:appx:magnus-expansion-term-tree-repr} can be rewritten using \cref{eq:appx:commutator-chain-separation,eq:tree-commutators-lie-algebra}, yielding the ME at order $k$ 
\begin{equation}
    M_k(t) = (-i)^{k-1} \sum_{\mu} \left[ \sum_{\tau\in \mathbb{T}_{k-1}} \alpha(\tau) \sum_{i=0}^{2^k-1} \beta_{\mu, i, \tau} \left( \fint_0^t dt_{[\tau]} \prod_{\{j \in [k]:i_j=1\}} d(t_j) \right) \right] L_\mu \label{eq:ME-before-tensor}
\end{equation}
Since we have written the control function 
\begin{equation}
    d(t) = \sum_{a=0}^{m} \frac{d_a}{a!} t^{a}
\end{equation}
as a polynomial, the integral term becomes
\begin{equation}
\begin{split}
    \fint_0^t dt_{[\tau]} \prod_{\{j\in[k]: i_j=1\}} d(t_j) &= \fint_0^t dt_{[\tau]} \prod_{\{j\in[k]:i_j=1\}} \left( \sum_{a_j=0}^m \frac{d_{a_j}}{a_j!} (t_j)^{a_j} \right) \\
    &= \sum_{\left\{\substack{a_j=0,\ldots,m \\ j\in[k], i_j=1} \right\} } \left( \prod_{\{j\in[k]:i_j=1\} }d_{a_j} \right) \left( \fint_0^t dt_{[\tau]} \prod_{\{j\in[k]:i_j=1\}} \frac{(t_j)^{a_j}}{a_j!} \right)
\end{split}
\end{equation}
where $a_j$ are the summation indices of the polynomial corresponding to $d(t_j)$. The product of sums
\begin{equation}
    \prod_{\{j\in[k]: i_j=1\}}\left(\sum_{a_j=0}^m f(a_j) \right)=\sum_{\left\{\substack{a_j=0,\ldots,m \\ j\in[k]: i_j=1} \right\} } \prod_{\{j\in[k]: i_j=1\} }f(a_j)  
\end{equation}
has been rewritten as a multi-dimensional sum. For example, if $i=101_2$, the sum becomes
\begin{equation}
    \sum_{ \left\{\substack{a_j=0, \ldots,m\\ j\in\{1,3\}}\right\} } \prod_{j\in\{1,3\}} f(a_j) = \sum_{a_1=0}^m\sum_{a_3=0}^m f(a_1)f(a_3).
\end{equation}
We can rearrange the sum in the $i$ and $a_j$ by grouping together the product of the polynomial coefficients $d_a$ with the same number of products between $d_{a_j}$. That is, the sum transforms
\begin{equation}
\begin{split}
    &\sum_{i=0}^{2^k-1}\sum_{\left\{\substack{a_j=0,\ldots,m \\ j\in[k], i_j=1} \right\} } b_i \left( \prod_{\{j\in[k]:i_j=1\}} f(a_j) \right) g(i,\{a_j:j\in[k], i_j=1\}) \\
    &= \sum_{p=0}^k \sum_{\gamma_1=0}^m\ldots\sum_{\gamma_p=0}^m\left[  \left(\prod_{p=1}^p f(\gamma_p)\right) \sum_{0<q_1<\dots<q_p\leq k} b_{q_1,\dots,q_p} g'(q_1, \ldots, q_p,\gamma_1,\ldots\gamma_p)\right].
\end{split}
\end{equation}
where we have the change of indices $\{q_1,\ldots,q_p\}=\{j\in[k]: i_j=1\}$, $\{\gamma_1,\ldots,\gamma_p\}=\{a_j: j\in[k], i_j=1\} $ and $g'(\cdot)$ returns the same as $g(\cdot)$ according to this index change
\begin{equation}
g'(q_1,\ldots,q_p,\gamma_1,\ldots,\gamma_p) := g\left(i(q_1,\ldots,q_p) ,\{\gamma_1,\ldots, \gamma_p\}\right).
\end{equation}
with $i(q_1,\ldots,q_p)$ written in the binary representation
\begin{equation}
    i(q_1,\ldots,q_p):=i_k\ldots i_1, \quad i_{j}(q_1,\ldots,q_p):=\begin{cases}
        1 & j\in \{q_1,\ldots,q_p\}\\
        0 & \text{otherwise}
    \end{cases}.
\end{equation}
In essence, these $q$'s indicate the position of the $1$'s in $i_{\text{base}\ 2}$. To give a simple example with $k=3$, this sum corresponds to the following grouping of the terms
\begin{equation}
\begin{split}
        &b_{000_2}\, g(0, \{\}) + b_{001_2} f(a_{1}) g(1, \{a_1\})  + b_{010_2} f(a_2) g(2, \{a_2\}) + b_{011_2} f(a_1) f(a_2) g(3, \{a_1, a_2\}) \\
        +& b_{100_2} f(a_3) g(4, \{a_3\})+ b_{101_2} f(a_1) f(a_3) g(5, \{a_1, a_3\}) + b_{110_2} f(a_2) f(a_3) g(6, \{a_2, a_3\}) \\
        +& b_{111_2} f(a_1) f(a_2) f(a_3) g(7, \{a_1,a_2,a_3\}) \\
        &=\Big[ b_{000_2}\, g(0, \{\}) \Big]\quad ({p=0}) \\
        &+ \Big[b_{001_2} f(a_{1}) g(1, \{a_1\})+ b_{010_2} f(a_2) g(2, \{a_2\}) + b_{100_2} f(a_3) g(4, \{a_3\})\Big]\quad ({p=1}) \\
        &+ \Big[ b_{011_2} f(a_1) f(a_2) g(3, \{a_1, a_2\}) + b_{101_2} f(a_1) f(a_3) g(5, \{a_1, a_3\})+ b_{110_2} f(a_2) f(a_3) g(6, \{a_2, a_3\})  \Big] \quad ({p=2})\\
        &+ \Big[b_{111_2} f(a_1) f(a_2) f(a_3) g(7, \{a_1,a_2,a_3\}) \Big], \quad (p=3)
\end{split}
\end{equation}
where the sums in $a_1,a_2,a_3$ have been neglected for simplicity. 
The inner sum in \cref{eq:ME-before-tensor} can then be rewritten as the following
\begin{equation}
\begin{split}
    &\sum_{i=0}^{2^k-1}\beta_{\mu, i, \tau}  \sum_{\left\{\substack{a_j=0,\ldots,m \\ j\in[k], i_j=1} \right\} } \left(\prod_{\{j\in[k]:i_j=1\}}d_{a_j}\right) \fint_0^t dt_{[\tau]} \prod^k_{\{j\in[k]:i_j=1\}} \frac{(t_j)^{a_j}}{a_j!}\\
    =&\sum_{p=0}^k \sum_{\gamma_1,\ldots,\gamma_p=0}^mt^{k}\left( \prod_{l=1}^p d_{\gamma_l} t^{\gamma_l}\right)  \left[\sum_{0<q_1<\cdots<q_p\leq k} \hspace{-4mm}\beta_{\mu, q_1 \cdots q_p, \tau}  \fint_0^1 dx_{[\tau]} \prod_{l=1}^p \frac{(x_{q_l})^{\gamma_l}}{\gamma_l!}\right],
\end{split}
\end{equation}
where we have applied the change of integration variable $dt_i = t\, dx_i$, $i=1,\ldots,k$. 
Notice the re-indexing from the binary representation of $i$ to the $q$'s. The term within the square brackets is a coefficient that depends only on $\mu,\gamma_1\cdots \gamma_p, \tau$ for a specific tree $\tau$, independent of the control function $d(t)$. Since $\beta_{\mu, \alpha_1 \cdots \alpha_p, \tau}$ is a coefficient in this sum, this object also includes the properties of the Lie algebra. Including now the sum over the trees of order $k$ results in the formula for the dynamical coefficient
\begin{equation}
    \mathbf{S}_{\mu, \vec \gamma }^{(k, p)} = \frac{(-i)^{k-1}}{\gamma_l!} \sum_{\tau\in \mathbb{T}_{k-1}} \alpha(\tau)\hspace{-5mm} \sum_{0<q_1 <\cdots<q_p\leq k}\hspace{-1mm} \left( \beta_{\mu, q_1 \cdots q_p, \tau} \fint_0^1\hspace{-2mm} dx_{[\tau]} \prod_{l=1}^p (x_{q_l})^{\gamma_l} \right).\label{eq:ME-tensor-calc} 
\end{equation}
The method to systematically solve the integral term is described in \cref{sect:appx-integral-over-tree}. The Magnus Expansion term of order $k$ then becomes 
\begin{equation}
    M_k(t, \vec d\,) = \sum_\mu\left[ \sum_{p=0}^k \sum_{\gamma_1 \cdots\gamma_p}\mathbf{S}_{\mu, \vec \gamma}^{(k,p)}\ t^k\left( \prod_{l=1}^p d_{\gamma_l} t^{\gamma_l}\right)  
    \right] L_\mu, \label{eq:ME-tensor-repr}
\end{equation}
where the term within square brackets is simply a coefficient for the Lie algebra element $L_\mu$. One can consider $t<1$ and neglect terms smaller $\mathcal{O}(t^{\Gamma})$, i.e. we truncate at chosen time order $\Gamma$. Thus, the tensor $\mathbf{S}_{\mu, \vec \gamma}^{(k,p)}$ is computed for all $\gamma$'s such that $k+\sum_{i=1}^p \gamma_i \leq \Gamma$. This is a reasonable approximation as long as $\Gamma> m, k_{\text{M}}$. Furthermore, since the evaluation of the polynomial is independent of the ordering of $d_{\gamma_i}$, the sum over $\gamma_1,\ldots,\gamma_p$ can be simplified to a sum where $\gamma_i\leq\gamma_{i+1}$ with $i=1,\ldots, p-1$, resulting in the dynamical coefficients $\mathbf{T}_{\mu, \vec \gamma}^{(k,p)}$ as described in the main text.




\subsection{Lie Algebra commutators over a tree}\label{sect:appx-lie-algebra-over-tree}
\begin{figure}
    \centering
    \includegraphics[width=0.7\linewidth]{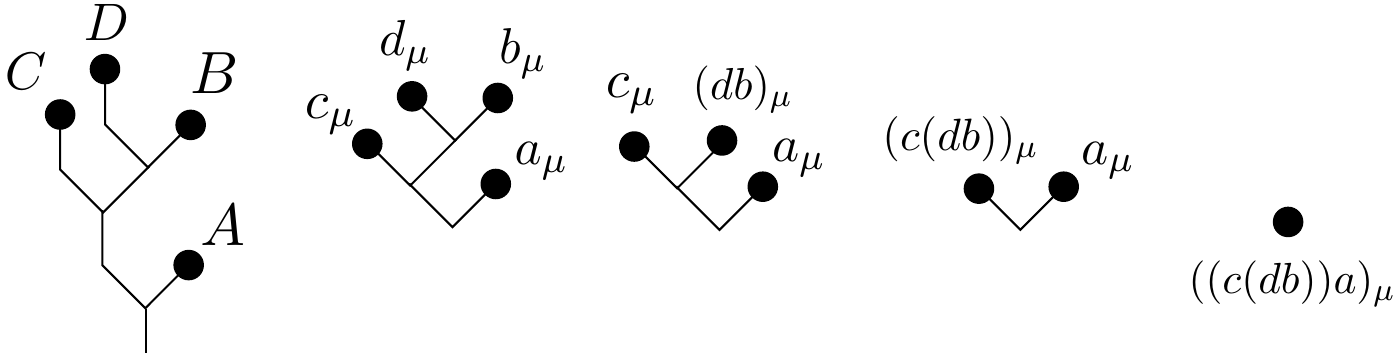}
    \caption{Example of evaluating a chained integral defined over a binary tree. The operators $A,B,C,D$ are placed at the leaves, which are represented in the Lie algebra with $ A=\sum_\mu a_\mu L_\mu$ (equivalently for $B, C, D$). We turn into the Lie algebra representation at the second step. Here we also remove vertical branches as they represent integration, which is unnecessary for time-independent operators. The commutation is evaluated in topological order, and occur between the members at each end of the branch. Since every operator here belongs to the Lie algebra, their transformation under commutation is easily described with the structure tensor. }
    \label{fig:binary-tree-commutators}
\end{figure}

In this section we want to define the identity \cref{eq:tree-commutators-lie-algebra}. After generating the independent Lie algebra and the corresponding structure tensor \cref{eq:independent-lie-algebra,eq:structure-tensor}, we can evaluate the recursive commutators defined over the trees in a similar fashion as with the chained integrals. Consider two operators $A,B$ which belong to the Lie Algebra~$\mathfrak{g}$
\begin{equation}
    A = \sum_{L_\mu \in \mathfrak{g}} a_\mu L_\mu, \ \
    B = \sum_{L_\mu \in \mathfrak{g}} b_\mu L_\mu.
\end{equation}
Their commutation is easily evaluated using the Lie algebra structure tensor
\begin{equation}
    [A,B] = \sum_{\mu,\nu} a_\mu b_\nu\, [L_\mu,L_\nu] = \sum_{\alpha} \left( \sum_{\mu, \nu} f_{\mu \nu \alpha} a_\mu b_\nu \right) L_\alpha = \sum_{\alpha}\, (ab)_\alpha\, L_\alpha 
\end{equation}
Since the result belongs to the same algebra, this becomes a simple tensor contraction when evaluated numerically. An example of evaluating the recursive commutators defined over a tree is shown in \cref{fig:binary-tree-commutators}.

\subsection{Integration of polynomials over a tree} \label{sect:appx-integral-over-tree}
\begin{figure}
    \centering
    \includegraphics[width=0.8\linewidth]{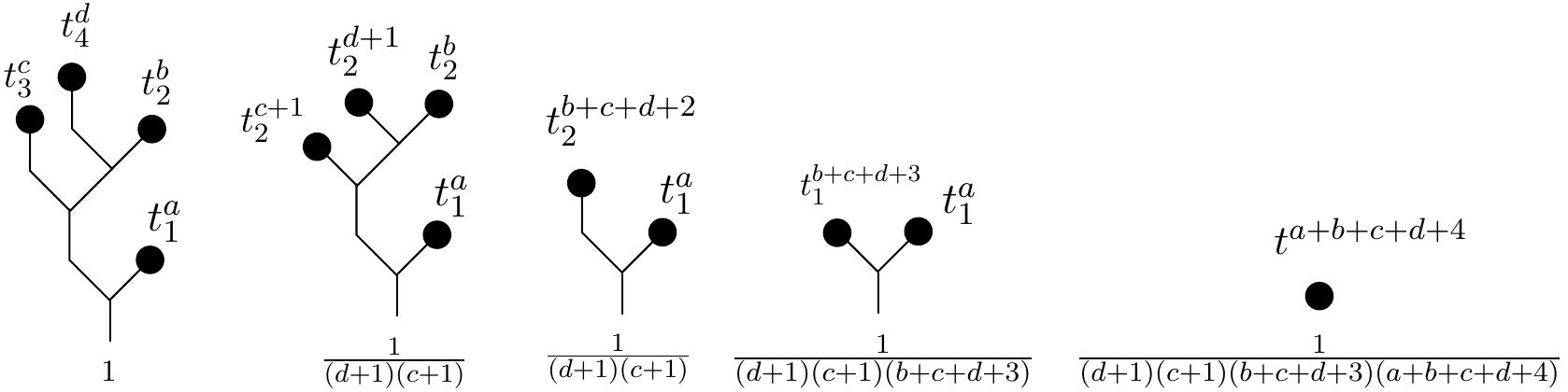}
    \caption{Example of solving a chained integral defined from a binary tree. Powers of the integration time are placed at the leaves, and these are worked out downwards, in topological order. In the first step, two integrals are resolved, introducing a coefficient, shown at the root of the tree. The powers are aggregated when there is a branch before computing the following integral. The end result is always $t$ to the power of the sum of the leaf powers plus the number of leaves (identical to the order of the tree) with a coefficient which depends on the topological transversal of the binary tree.}
    \label{fig:binary-tree-integral}
\end{figure}

In this section we describe how the polynomial integrals 
\begin{equation}
    \fint_0^t\hspace{-2mm} dt_{[\tau]} \prod_{l=1}^p (t_{q_l})^{\gamma_l} 
\end{equation}
can be computed systematically. 
When calculating a chained integral defined with a tree, we need to label each of the leaves of the tree, from $1$ to $k$. Then the indices $q_1$ to $q_p$ define where the powers $(t_{q_l})^{\gamma_l}$ go on the leaves of the tree. To solve the integral, we start from the leaves of the tree and work downwards, in topological order, until the root of the tree is reached. Going down a vertical branch corresponds to evaluating the integral (see example in \cref{fig:binary-tree-integral}). Consider that $q$ is a label of a leaf which is connected to a vertical branch, and $p$ is the label of the leaf at the root of the vertical branch. Then the integral can be directly evaluated
\begin{equation}
    \int_0^{t_p}\hspace{-2mm} dt_q\ (t_q)^{\gamma_q} =\frac{1}{\gamma_q+1} (t_p)^{\gamma_q+1}
\end{equation}
This term would be embedded in the integral of leaf $p$ 
\begin{equation}
  \frac{1}{\gamma_q+1} \int_0^{t_a} dt_p\ (t_p)^{\gamma_q+1}  (t_p)^{\alpha}/\beta
\end{equation}
 where $a$ is the label of the leaf at the root of vertical branch $p$,  $\alpha$ is an integer and $\beta$ a real coefficient computed from other branches. It is apparent that this new integral can be evaluated in a similar fashion, incrementing the power and updating the coefficient. 



\section{Optimal control of continuous functions} \label{appx:sect:optimal-control}

\subsection*{Control derivatives of the Magnus Expansion}

The expressions for the derivatives are polynomials represented as tensor contractions, and are easily obtained from the original set of coefficients $\mathbf{T}_{\mu,\vec \gamma}^{(k,p)}$. The time derivative of $a_\mu$ is given by
\begin{equation}
    \frac{\partial a_\mu}{\partial t} = \sum_{k=1}^{k_{\text{M}}} \sum_{\substack{p=0}}^k \sum_{\gamma_1, \ldots,\gamma_p}  \mathbf{\widetilde T}_{\mu,\vec \gamma}^{(k,p)} t^{k-1} \prod_{q=1}^p  d_{\gamma_q}t^{\gamma_q} ,\label{eq:ME-diff-time}
\end{equation}
where
\begin{equation}
   \mathbf{\widetilde T}_{\mu,\vec \gamma}^{(k,p)} := 
       ({ k+\textstyle \sum_q\gamma_q})  \ \mathbf{T}_{\mu,\vec \gamma}^{(k,p)}.
\end{equation}
The derivative with respect to the control polynomial coefficients $\vec d$ is
\begin{equation}
\frac{\partial a_\mu}{\partial d_\alpha} = \sum_{k=1}^{k_{\text{M}}} \sum_{\substack{p=1}}^k\sum_{\gamma_1, \ldots,\gamma_{p-1}}  \mathbf{D}_{\mu,\vec \gamma}^{(\alpha,k,p)} t^{k+\alpha} \prod_{q=1}^{p-1} d_{\gamma_q} t^{\gamma_q}, \label{eq:ME-diff-controls}
\end{equation}
where 
\begin{equation}
    \mathbf{D}_{\mu,\vec \gamma}^{(\alpha,k,p)}  := \, \mathbf{T}_{\mu,\ (\alpha, \gamma_1 ,\ldots,\gamma_{p-1})}^{(k,p)}+\mathbf{T}_{\mu,\ (\gamma_1 ,\alpha, \gamma_2,\ldots,\gamma_{p-1})}^{(k,p)} +\cdots  + \mathbf{T}_{\mu,\ (\gamma_1, \ldots,\gamma_{p-1}, \, \alpha)}^{(k,p)}
\end{equation}
are the coefficients for the control derivative of the dynamical coefficients.
\subsection*{Pulse sequence derivatives}
The pulse sequence  $U(\vec c)$ with the set of controls $\vec c$ is composed of $S$ segments
\begin{equation}
    U(\vec c) = \prod_{s=1}^S U_s(\vec c),
\end{equation}
where $U_s$ is the propagator at segment $s$ and $\vec c$ is a set of tunable parameters. The gradients of $U_s(\vec c)$ are easy to evaluate when these are generated by Pauli operators. When the propagator $U_s(\vec c)$ is generated by a dynamical evolution with singular control \cref{eq:hamiltonian-single-control}, we can use the derivatives of the ME we obtained above \cref{eq:ME-diff-time,eq:ME-diff-controls}. More specifically, consider that $U_s$ is generated from the evolution of a time-independent operator, similar to an effective Hamiltonian
\begin{equation}
    U_s(\vec c_s) = e^{-i H_s(\vec c_s) },
\end{equation}
where its derivative is
\begin{equation}
    \partial_{\vec c_s} U_s = -i U_s(\vec c_s) \sum_{k=0}^{k_D\rightarrow\infty}\frac{i^k}{(k+1)!} \text{ad}^k_H (\partial_{\vec c_s} H_s ),
\end{equation}
with the adjoint action $\text{ad}^{1}_A (B)=[A,B]$ and $\text{ad}^{k+1}_A (B)=[A,\text{ad}^k_A(B)]$, for any operators $A,B$. This effective Hamiltonian $H_s(\vec c_s)$ can be approximated by the truncated ME when $t<t_*$, and so $H\simeq M^{(k_{\text{M}})}=:M$. We consider that the derivative is accurately approximated when $k_D\ge k_{\text{M}}$, and thus choose $k_D = k_{\text{M}}$. The derivatives $\partial_{\vec c} M$ have been obtained in \cref{eq:ME-diff-time,eq:ME-diff-controls} when $\vec c=(t, d_0, \ldots,d_m)$ describes a single control function $d(t)$ expressed as a polynomial in time. Since $\partial_{\vec c} M$ are expressed in terms of the Lie-algebra elements, evaluating the adjoint action can be done using only the structure constant $f_{ijk}$ instead of the full Hilbert space. 


\subsection*{Transformation to the Hermite spline}
The control pulse $d(t)$ to be designed is manipulated using Hermite spline nodes.  Consider we have $S$ spline segments. Then, there are $S+1$ nodes placed at times $(0, t_1,\ldots,t_S)$, where $t_S$ is the total duration of the pulse. Each node 
\begin{equation}
    h^{(s)}(t-t_s) = \sum_{l=0}^{L} h_l^{(s)} \frac{(t-t_s)^l}{l!} 
\end{equation}
is a polynomial that matches the value and $L$ first derivatives of $d(t_s)$ 
\begin{equation}
    h_l^{(s)} = \left. \frac{d^l}{dt^l} d(t) \right|_{t=t_s}. \qquad (l=0,\ldots, L) \label{eq:hermite-spline-nodes}
\end{equation}
The shape of the control pulse is fully parametrized with 
\begin{equation}
    \vec h=(\Delta t_1',\ldots,\Delta t_S', h_0^{(0)}, \ldots, h_L^{(S)} ). \label{eq:hermite-parameter-transform}
\end{equation}
The segment duration is tuned with the time intervals $\Delta t_s'=t_s-t_{s-1},\ s=1,\ldots, S$. To each of these segments, indexed by $s=1,\ldots, S$, there is a polynomial
\begin{equation}
    d_s(t) = \sum_{n=0}^m d_n^{(s)} \frac{t^n}{n!}, \quad  t\in[0, \Delta t_s]
\end{equation}
that matches its derivatives at the past node  $h_{s-1}(t)$ and the derivatives at future node $h_s(t)$, where $m=2L+1$. This leads to the constraints 
\begin{align}
    \Delta t_s' &=\Delta t_s, \qquad (s=1,\ldots,S), \\ 
    h_l^{(s-1)} &= d_l^{(s)} , \qquad (s=1,\ldots,S), \ (l = 0, \ldots, L)\label{eq:hermite-constraints1}\\
    h_l^{(s)} &=\sum_{n=l}^m d_{n}^{(s)} \frac{(\Delta t_s)^{n-l}}{(n-l)!} ,  \qquad (s=1,\ldots,S), \ (l = 0, \ldots, L) \label{eq:hermite-constraints2} 
\end{align}
where \cref{eq:hermite-constraints1,eq:hermite-constraints2} are solved by constructing a linear system of equations. 
We can evaluate the ME and its derivatives using the segment duration and polynomial coefficients at each segment $\vec c_s(\vec h)=(\Delta t_s, d_0^{(s)}, \ldots, d_m^{(s)})(\vec h)$.
These must be transformed into the polynomial coefficients $\vec c(\vec h)=(\vec c_1, \ldots, \vec c_S)(\vec h)$ so that $M$ and $\partial_{\vec c} M$ can be evaluated at each segment. This is a change of coordinates $\vec c(\vec h)$, where the Jacobian of the transformation is $\partial \vec c/\partial \vec h$ and is obtained from the constraints in \cref{eq:hermite-constraints1,eq:hermite-constraints2}, such that the derivatives $\partial_{\vec c}U$ 
\begin{equation}
    \delta U = \left(\frac{\partial U}{\partial \vec c}\right)^{T}  \cdot \delta \vec c = (\delta \vec c)^{T} \cdot \left( \frac{\partial U}{\partial \vec c}\right)
\end{equation}
are transformed into $\partial_{\vec h}U$. Since $\delta \vec c = (\partial\vec c/\partial \vec h) \cdot \delta \vec h$, we get
\begin{equation}
    \delta U = (\delta \vec h)^T\cdot \left(\frac{\partial \vec c }{\partial \vec h}\right)^T \cdot  \left( \frac{\partial U}{\partial \vec c} \right) \Leftrightarrow \frac{\partial U}{\partial \vec h} = \left(\frac{\partial \vec c }{\partial \vec h}\right)^T \left(\frac{\partial U}{\partial \vec c}\right). 
\end{equation}
The variational of the constraint equations  \begin{align}
    \delta (\Delta t_s') &= \delta(\Delta t_s), \qquad (s=1,\ldots,S)\\
    \delta h_s^{(s-1)} &= \delta d_l^{(s)}, \qquad (s=1,\ldots,S), \ (l = 0, \ldots, L)\\
    \delta h_l^{(s)} &= \sum_{n=l}^{m}\frac{(\Delta t_s)^{n-l}}{(n-l)!}\delta d_n^{(s)}  + \left(\sum_{n=l+1}^{m} d_{n}^{(s)} \frac{(\Delta t_s)^{n-l-1}}{(n-l-1)!} \right) \delta(\Delta t_s), \qquad (s=1,\ldots,S), \ (l = 0, \ldots, L)
\end{align}
is used for the calculation of the Jacobian $\partial \vec c/\partial \vec h$.
This set of equations can be written as $A\, \delta \vec h = B \, \delta \vec c$ with $A,B$ matrices obtained from these equations. Then, the Jacobian is simply
\begin{equation}
    \frac{\partial \vec c}{\partial \vec h} = B^{-1} A,
\end{equation}
where $B^{-1}$ is the pseudo-inverse of $B$. 

\end{document}